\documentclass{amsart}

\RequirePackage{amsthm,amsmath,amsfonts,amssymb}
\RequirePackage[authoryear]{natbib}
\RequirePackage[colorlinks,citecolor=blue,urlcolor=blue]{hyperref}
\RequirePackage{graphicx}

\usepackage{tikz}
\usepackage{pifont} 
\usepackage{multirow}
\usepackage{multicol}
\usepackage{colortbl} 
\usepackage{amsaddr}
\usepackage{lmodern}
\usepackage{fancyhdr}
\makeatletter
\newcommand{\thickhline}{%
    \noalign {\ifnum 0=`}\fi \hrule height 1pt
    \futurelet \reserved@a \@xhline
}
\newcolumntype{"}{@{\hskip\tabcolsep\vrule width 1pt\hskip\tabcolsep}}
\makeatother

\begin{document}

\title{Clustering of football players based on performance data and aggregated clustering validity indexes}

\author{Serhat Emre Akhanlı}
\address{Muğla Sıtkı Koçman University, Department of Statistics, Muğla, Turkey}
\email{serhatakhanli@mu.edu.tr}
\author{Christian Hennig}
\address{University of Bologna, Department of Statistical Sciences ``Paolo Fortunati'', via delle Belle Arti, Bologna, Italy}
\email{christian.hennig@unibo.it}

\maketitle
\begin{abstract}
We analyse football (soccer) player performance data with mixed type variables from the 2014-15 season of eight European major leagues. We cluster these data based on a tailor-made dissimilarity measure. 

In order to decide between the many available clustering methods and to choose an appropriate number of clusters, we use the approach by \cite{akhanli2020comparing}. This is based on several validation criteria that refer to different desirable characteristics of a clustering. These characteristics are chosen based on the aim of clustering, and this allows to define a suitable validation index as weighted average of calibrated individual indexes measuring the desirable features.

We derive two different clusterings. The first one is a partition of the data set into major groups of essentially different players, which can be used for the analysis of a team's composition. The second one divides the data set into many small clusters (with 10 players on average), which can be used for finding players with a very similar profile to a given player. It is discussed in depth what characteristics are desirable for these clusterings. Weighting the criteria for the second clustering is informed by a survey of football experts.
\end{abstract}




\section{Introduction}
\label{sec:intro} 

Nowadays, a large amount of performance data of professional football (soccer) players is routinely collected. The analysis of such data is of great commercial interest. Here we cluster complex player performance data with mixed type variables from the 2014-15 season of eight European major leagues. 

Sports have embraced statistics in assisting player recruitment and playing strategies. Different statistical methodologies have been applied to various types of sports data. Cluster analysis has been used for aggregating similar types of players in several applications. \cite{ogles2003typology} suggested that by using cluster analysis (Ward's method), marathon runners can be categorised into five groups in terms of their motives for running. \cite{gaudreau2004different} examined coping strategies used by groups of athletes based on a hierarchical cluster analysis using Ward's method. \cite{wang2009intra} observed coaching behaviour among basketball players, and showed that three distinct groups could be identified by using an agglomerative hierarchical clustering method. \cite{Yingying2010ModelingAP} applied different clustering techniques to athletes' physiological data, and proposed a new hierarchical clustering approach. \cite{Kosmidis2015} used NBA players’ data to form groups of players in terms of their performance using copula-based finite mixture models. \cite{DuttaYurkoVentura} adopted model based clustering for data of defensive NFL players.  

There is also connected work on football data. \cite{bialkowski2014identifying} adopted k-means clustering and minimum entropy data partitioning to identify a team's structure. \cite{feuerhake2016recognition} used the Levenstein distance and then k-means and DBSCAN clustering to analyse sequences of movements in a soccer game. \cite{hobbs2018quantifying} applied spatio-temporal trajectory clustering that could automatically identify counter-attacks and counter-pressing without requiring unreliable human annotations. \cite{decroos2020player} created a ``player vector'' that characterizes a player's playing style using methods such as clustering and nearest neighbour. 

A key contribution of the present work is the assessment of the quality of different clusterings, which allows us to select from a wide range of clustering solutions for the analysed data set coming from different clustering approaches and numbers of clusters. \cite{hennig2015true,hennig2015clustering31} have argued that there is no single ``true'' clustering for a given data set, and that the quality of different clusterings depends on the requirements of the specific application, and in particular on what characteristics make a clustering desirable for how the clusters are later used and interpreted. Different uses can be imagined for clusterings of football players according to performance data, and we aim at measuring clustering quality with such uses in mind. We propose two different such measurements for different aims of clustering. The first one is to give a rough representation of the structure in the data in terms of a low number of clusters corresponding to easily interpretable types of players. This can be used for example to analyse team compositions and positioning in terms of these clusters, and to relate it to success. The second one is to have small clusters of very similar players that can be used for finding potential replacements for a player, and to analyse similarities between teams on a finer scale. The second aim requires a much larger number of clusters than the first one. Arguably, none of the existing standard methods for determining the number of clusters in the literature (see Section \ref{sec:indexes}) is reliable when comparing very small (around 4, say) with very large (more than 100) numbers of clusters based on the data alone. In fact, on most data sets, these will not directly compete. Rather it depends on the clustering aim whether a rather small or a rather large number of clusters is required. 

We will take the approach proposed by \cite{hennig2017cluster} and elaborated in \cite{akhanli2020comparing}, which is based on a set of indexes that are meant to measure different desirable features of a clustering in a separate manner, and then the user can select indexes and weights according to the requirements of the application in order to define a composite index. This requires a calibration scheme that makes the values of the different indexes comparable, so that their weights can be interpreted in terms of the relative importance of the respective characteristic. Although we analyse data from the 2014-15 season, the composite indexes resulting from this approach are applicable to other data sets of a similar kind. 

Another important ingredient of our clusterings is a suitable dissimilarity measure between players. This involves a number of nontrivial choices, as the data are of mixed type (there are categorical position variables, counts, ratios, and compositional variables as well as variables that are very skewly distributed and require transformation and other ways of re-expression). A suitable dissimilarity measure for football player performance data was proposed in \cite{akhanli2017some} with the intention to use it for mapping the players by means of multidimensional scaling (MDS) \citep{BorGro12} and dissimilarity-based clustering. Some details that were not covered in \cite{akhanli2017some} are explained here.

In Section~\ref{sec:datadissimilarity} the data set is introduced and the dissimilarity measure is defined. Section~\ref{sec:cmethods} lists the cluster analysis methods that have been used. Section~\ref{sec:aggregation_indexes} introduces various indexes for cluster validation from the literature, and the indexes used for individual aspects of clustering quality along with the calibration and weighting scheme according to \cite{akhanli2020comparing}. Section~\ref{sec:valresults} applies these ideas to the football players data set. This includes a discussion of the weights to be chosen, which involves a survey among football experts regarding whether specific players should be clustered together in order to justify one of the weighting schemes. Section~\ref{sec:conclusion} concludes the paper.  

\subsection{General notation} 
\label{sec:general_notation}

Given a data set, i.a., a set of distinguishable objects $\mathcal{X}=\left\{ x_{1}, x_{2}, \ldots, x_{n} \right\}$, the aim of cluster analysis is to group them into subsets of $\mathcal{X}$. A clustering is denoted by $\mathcal{C}=\left\{ C_{1}, C_{2}, \ldots, C_{K} \right\}$, $C_k\subseteq \mathcal{X},$ with cluster size $n_{k}= |C_{k}|,\ k=1,\ldots,K$. We require $\mathcal{C}$ to be a partition, e.g., $k \neq g  \Rightarrow C_{k} \cap C_{g} = \emptyset$ and $\bigcup_{k=1}^{K} C_{k} = \mathcal{X}$. Clusters are assumed to be crisp rather than fuzzy, i.e., an object is either a full member of a cluster or not a member of this cluster at all. An alternative way to write $x_{i}\in C_k$ is $l_i=k$, i.e., $l_i\in\{1,\ldots,K\}$ is the cluster label of $x_{i}$.

The approach presented here is defined for general dissimilarity data. A dissimilarity is a function $d : \mathcal{X}^{2} \rightarrow \mathbb{R}_{0}^{+}$ so that $d(x_{i}, x_{j}) = d(x_{j}, x_{i}) \geq 0$ and $d(x_{i}, x_{i}) = 0$ for $x_{i}, x_{j} \in \mathcal{X}$. Many dissimilarities are distances, i.e., they also fulfill the triangle inequality, but this is not necessarily required here. 

\section{Football players dataset and dissimilarity construction}
\label{sec:datadissimilarity}

The data set analysed here contains 1501 football players characterized by 107 variables. It was obtained from the website \url{www.whoscored.com}. Data refer to the 2014-2015 football season in 8 major leagues (England, Spain, Italy, Germany, France, Russia, Netherlands, Turkey). The original data set had 3003 players, which were those who have appeared in at least one game during the season. Goalkeepers have completely different characteristics from outfield players and were therefore excluded from the analysis. Because data about players who did not play very often are less reliable, and because the methods that we apply are computer intensive, we analysed the 1501 (about 50\%) players who played most (at least 1403 or 37\% out of a maximum of 3711 minutes). Variables are of mixed type, containing binary, count and continuous information. The variables can be grouped as follows:

\begin{itemize}
	\item \textbf{Team and league variables}: League and team ranking score based on the information on UEFA website, and team points from the ranking table of each league,
	\item \textbf{Position variables}: 11 variables indicating possible positions on which a player can play and has played,
	\item \textbf{Characteristic variables}:  Age, height, weight,
	\item \textbf{Appearance variables}:  Number of appearances of teams and players, and players number of minutes played, 
	\item \textbf{Top level count variables}: Interceptions, fouls, offsides, clearances, unsuccessful touch, dispossess, cards, etc.
	\item \textbf{Lower level count variables}: Subdivision of some top level count variables as shown in Table~\ref{tab:lowerlevel} 
\end{itemize}

\begin{table}[h]
	\renewcommand{\arraystretch}{1.25}

	\caption{Top and lower level count variables \label{tab:lowerlevel}}
	\begin{tabular}{p{2cm} p{9.5cm}}
		\thickhline
		\textit{TOP LEVEL} & \textit{LOWER LEVEL} \\
		\thickhline
		\rowcolor{lightgray}
		&\textit{Zone:} Out of box, six yard box, penalty area \\ 
		\rowcolor{lightgray}
		&\textit{Situation:} Open play, counter, set piece, penalty taken \\
		\rowcolor{lightgray}
		&\textit{Body part:} Left foot, right foot, header, other \\
		\rowcolor{lightgray}
		\multirow{-4}{*}{\textbf{SHOT}} &\textit{Accuracy:} On target, off target, blocked \\

		\multirow{3}{*}{\textbf{GOAL}}
		& \textit{Zone:} Out of box, six yard box, penalty area \\
		& \textit{Situation:} Open play, counter, set piece, penalty taken \\
		& \textit{Body part:} Left foot, right foot, header, other \\

		\rowcolor{lightgray}
		& \textit{Length:} AccLP, InAccLP, AccSP, InAccSP \\
		\rowcolor{lightgray}
		\multirow{-2}{*}{\textbf{PASS}} & \textit{Type:} AccCr, InAccCr, AccCrn, InAccCrn, AccFrk, InAccFrk \\

		\multirow{2}{*}{\textbf{KEY PASS}}
		& \textit{Length:} Long, short \\
		& \textit{Type:} Cross, corner, free kick, through ball, throw-in, other  \\

		\rowcolor{lightgray}
		\textbf{ASSIST} & \textit{Type:} Cross, corner, free kick, through ball, throw-in, other  \\

		\textbf{BLOCK} & Pass blocked, cross blocked, shot blocked  \\
		\rowcolor{lightgray}
		\textbf{TACKLE} & Tackles, dribble past  \\
		\textbf{AERIAL} & Aerial won, aerial lost  \\
		\rowcolor{lightgray}
		\textbf{DRIBBLE} & Dribble won, dribble lost  \\
		\thickhline
		\multicolumn{2}{l}{\scriptsize{*Acc: Accurate, *InAcc: Inaccurate}} \\ 
		\multicolumn{2}{l}{\scriptsize{*LP: Long pass, *SP: Short pass, *Cr: Cross, *Crn: Corner, *Frk: Free kick}}\\
	\end{tabular}
	\label{tab:data}
\end{table}

In order to appropriately take into account the information content in the different variables, \cite{akhanli2017some} constructed a dissimilarity measure between players, which we review here (the choice of $c$ in Section \ref{subsec:trans} was not explained there). See that paper for more details including missing value treatment. The construction process had five stages:
\begin{enumerate}
	\item \textbf{Representation:} Re-defining variables in order to represent the relevant information in the variables appropriately; 
	\item \textbf{transformation} of variables, where the impact of variables on the resulting dissimilarity is appropriately formalised in a nonlinear manner;
	\item \textbf{standardisation} in order to make within-variable variations comparable between variables;
	\item \textbf{weighting} to take into account that not all variables have the same importance;
	\item \textbf{aggregation:} Defining a dissimilarity putting together the information from the different variables; the first four stages need to be informed by the method of aggregation. 
\end{enumerate}

Data should be processed in such a way that the resulting dissimilarity between observations matches how dissimilarity is interpreted in the application of interest, see \cite{HenHau06,hennig2015clustering31}. The resulting dissimilarities between observations may strongly depend on transformation, standardisation, etc., which makes variable pre-processing very important. 

\subsection{Representation}
\label{subsec:repre}
Counts of actions such as shots, blocks etc. should be used relative to the period of time the player played. A game of football lasts for 90 minutes, so we represent the counts as ``per 90 minutes'', i.e., divided by the minutes played and multiplied by 90. We will still refer to these variables as ``count variables'' despite them technically not being counts anymore in this way. 

Regarding count variables at different levels such as shots overall, shots per zone, shot accuracy, there is essentially different information in (a) the overall number and (b) the distribution over sub-categories. Therefore the top level counts are kept (per 90 minutes), whereas the lower level counts are expressed as proportions of the overall counts. Some counts in sub-categories can be interpreted as successes of actions counted by other variables. For example there is accuracy information for passes, and goals are successful shots. In these cases, success rates are used (i.e., goals from the six yard box are expressed as success percentage of shots from the six yard box). In some cases both success rates and sub-category proportions are of interest in their own right, in which case they are both kept, see Table \ref{tab:repre} for an overview. Note that later variables are aggregated in such a way that redundant information (such as keeping all sub-category proportions despite them adding up to 1 and therefore losing a degree of freedom) does not cause mathematical problems, although this should be taken into account when weighting the variables, see Section \ref{subsec:weight}. 

\begin{table}[h]
	\renewcommand{\arraystretch}{1.25}
	\caption{Representation of lower level count variables \label{tab:rep} }
	\begin{tabular}{p{3.5cm}p{2.5cm}p{5.5cm}}
		\hline
		\textbf{Variables} & \textbf{Proportional total} & \textbf{Success rate} \\
		\textbf{(Include sub-categories)} & \textbf{(standardised by)} & \textbf{(standardised by)} \\
		\hline
		Block & Total Blocks & \ding{56} \\
		Tackle, Aerial, Dribble & \ding{56} & Total tackles, total aerials, and total dribbles \\
		Shot (4 sub-categories) & Total shots & \ding{56} \\
		Goal (4 sub-categories) & Total goals & Shot count in different sub-categories, and total shots for overall success rate \\
		Pass (2 sub-categories) & Total passes &  Pass count in different sub-categories, and total passes for overall success rate \\
		Key pass (2 sub-categories) & Total key passes &  \ding{56} \\
		Assist & Total assists & Key pass count in different sub-categories, and total assists for overall success rate \\
		\hline
	\end{tabular}
	\label{tab:repre}
\end{table}

\subsection{Transformation}
\label{subsec:trans}

The top level count variables have more or less skew distributions; for example, many players, particularly defenders, shoot very rarely during a game, and a few forward players may be responsible for the majority of shots. On the other hand, most blocks come from a few defenders, whereas most players block rarely. 
This means that there may be large absolute differences between players that shoot or block often, whereas differences at the low end will be low; but from the point of view of interpretation, the dissimilarity between two players with large but fairly different numbers of blocks and shots is not that large, compared with the difference between, for example, a player who never shoots and one who occasionally but rarely shoots. Most of these variables $x$ have therefore been transformed by $y=\log(x+c)$, where the constant $c$ (or no transformation) has been chosen dependently of the variable in question by taking into account data from the previous season. The transformation was chosen in order to make the differences between the two years as stable as possible over the range of $x$, according to the rationale that in this way the amount of ``random variation'' is near constant everywhere on the value range. More precisely, a regression was run, where the response was the absolute value of the player-wise transformed count difference between the two seasons, and the explanatory variable was the weighted mean (by minutes played) of the two transformed count values. $c$ is then chosen so that the regression slope is as close to zero as possible (see \cite{akhanlithesis} for more details and issues regarding matching player data from the two seasons). 

\subsection{Standardisation}
\label{subsec:stand}

The general principle of aggregation of variables will be to sum up weighted variable-wise dissimilarities (see Section \ref{sec:aggre}), which for standard continuous variables amounts to computing the $L_1$ (Manhattan) distance. Accordingly, variables are standardised by the average absolute distance from the median. For the lower level percentages, we standardise by dividing by the pooled average $L_1$ distance from the median. We pool this over all categories belonging to the same composition of lower level variables. This means that all category variables of the same composition are standardised by the same value, regardless of their individual relative variances. The reason for this is that a certain difference in percentages between two players has comparable meaning between the categories, which does not depend on the individual variance of the category variable (see \cite{akhanli2017some} for a discussion of the treatment of compositional variables). 

\subsection{Weighting}
\label{subsec:weight}
An aspect of variable weighting here is that in case that there are one or more lower level compositions of a top level variable, the top level variable is transformed and standardised individually, whereas the categories of the lower level percentage composition are standardised together. This reflects the fact that the top level count and the lower level distribution represent distinct aspects of a player's characteristics, and on this basis we assign the same weight to the top level variable as to the whole vector of compositional variables, e.g., a weight of one for transformed shot counts is matched by a weight of $1/3$ for each of the zone variables ``out of the box'', ``six yard box'', ``penalty area''. Implicitly this deals with the linear dependence of these variables (as they add to one); their overall weight is fixed and would not change if the information were represented by fewer variables removing linear dependence. 

In case that a top level count variable is zero for a player, the percentage variables are missing. In this situation, for overall dissimilarity computation between such a player and another player, the composition variables are assigned weight zero and the weight that is normally on a top level variable and its low level variables combined is assigned to the top level variable.

\subsection{Aggregation of variables}
\label{sec:aggre}
There are different types of variables in this data set which we treat as different groups of variables. There are therefore two levels of aggregation, namely aggregation within a group, and aggregation of the groups. Group-wise dissimilarities $d_k$ are aggregated as follows:

	\begin{equation}
		\label{eq:dist_agg}
			d_{fin}(\mathbf{x}, \mathbf{y}) = \sum_{k=1}^{3} \frac{w_{k}*d_{k}(\mathbf{x}, \mathbf{y})}{s_{k}},
	\end{equation}
	
\noindent where $w_{k}$ is the weight of group $k$, and $s_{k}$ is the standard deviation of the vector of all dissimilarities $d_k$ from group $k$. $w_k$ is chosen proportionally to the number of variables in the $k^{th}$ group. Note that there is another layer of weighting and standardising here on top of what was discussed in Sections \ref{subsec:stand} and \ref{subsec:weight}. This was done in order to allow for a clear interpretation of weights and measures of variability; it would have been much more difficult to standardise and weight individual variables of different types against each other. (\ref{eq:dist_agg}) takes inspiration from the Gower coefficient for mixed type data \citep{Gow71}, although Gower did not treat groups of variables and advocated range standardisation, which may be too dominated by outliers.
 
For quantitative variables (characteristics, appearances, top and lower level count variables), (\ref{eq:dist_agg}) with $d_k$ chosen as absolute value of the differences amounts to the $L_1$ (Manhattan) distance. These variables therefore do not have to be grouped. 

The league ranking scores and the team points from the ranking table of each league based on the 2014-2015 football season are aggregated to a single joint dissimilarity by adding standardised differences on both variables in such a way that a top team in a lower rated league is treated as similar to a lower ranked team in a higher rated league.   

The position variables can take values 0 or 1 for the presence, over the season, of the player on 11 different possible positions on the pitch. These are aggregated to a single dissimilarity using the geco coefficient for presence-absence data with geographical location, taking into account geographical distances, as proposed in \cite{HenHau06}, using a suitable standardised Euclidean distance between positions, see Table \ref{tab:dist_pos2}. 

\begin{table}
\scriptsize
	\centering
	\caption{Distances between each position. Here the values are obtained by using Euclidean geometry}
	\begin{tabular}{l|ccc c ccc ccc c}
		$\mathbf{d_{R}(a,b)}$ & DC & DL & DR & DMC & MC & ML & MR & AMC & AML & AMR & FW \\
		\hline
		DC & $0$ & $1$ & $1$ & $1$ & $2$ & $\sqrt{5}$ & $\sqrt{5}$ & $3$ & $\sqrt{10}$ & $\sqrt{10}$ & $4$  \\
		DL & $1$ & $0$ & $1$ & $\sqrt{2}$ & $\sqrt{5}$ & $2$ & $\sqrt{5}$ & $\sqrt{10}$ & $3$ & $\sqrt{10}$ & $\sqrt{17}$\\
		DR & $1$ & $1$ & $0$ & $\sqrt{2}$ & $\sqrt{5}$ & $\sqrt{5}$ & $2$ & $\sqrt{10}$ & $\sqrt{10}$ & $3$ & $\sqrt{17}$\\
		DMC & $1$ & $\sqrt{2}$ & $\sqrt{2}$ & $0$ & $1$	& $\sqrt{2}$ & $\sqrt{2}$ & $2$ & $\sqrt{5}$ & $\sqrt{5}$ & $3$ \\
		MC & $2$ & $\sqrt{5}$ & $\sqrt{5}$ & $1$ & $0$ & $1$ & $1$ & $1$ & $\sqrt{2}$ & $\sqrt{2}$ & $2$ \\
		ML & $\sqrt{5}$ & $2$ & $\sqrt{5}$ & $\sqrt{2}$ & $1$ & $0$ & $1$ & $\sqrt{2}$ & $1$ & $\sqrt{2}$ & $\sqrt{5}$ \\
		MR & $\sqrt{5}$ & $\sqrt{5}$ & $2$ & $\sqrt{2}$ & $1$ & $1$ & $0$ & $\sqrt{2}$ & $\sqrt{2}$ & $1$ & $\sqrt{5}$ \\
		AMC	& $3$ & $\sqrt{10}$	& $\sqrt{10}$ & $2$ & $1$ & $\sqrt{2}$ & $\sqrt{2}$ & $0$ & $1$ & $1$ & $1$ \\
		AML & $\sqrt{10}$ & $3$	& $\sqrt{10}$ & $\sqrt{5}$ & $\sqrt{2}$ & $1$ & $\sqrt{2}$ & $1$ & $0$ & $1$ & $\sqrt{2}$\\
		AMR	& $\sqrt{10}$ & $\sqrt{10}$	& $3$ & $\sqrt{5}$ & $\sqrt{2}$ & $\sqrt{2}$ & $1$ & $1$ & $1$ & $0$ & $\sqrt{2}$\\
		FW 	& $4$ & $\sqrt{17}$	& $\sqrt{17}$ & $3$ & $2$ & $\sqrt{5}$ & $\sqrt{5}$ & $1$ & $\sqrt{2}$ & $\sqrt{2}$ & $0$\\	
	\end{tabular}
	\label{tab:dist_pos2}
\end{table}

\section{Clustering methods}
\label{sec:cmethods}
Clustering has been carried out by standard dissimilarity-based clustering methods with the aim of finding the best clusterings by comparing all clusterings using a composite cluster validity index based on indexes measuring different aspects of clustering, see Section \ref{sec:aggregation_indexes}.

The following six clustering algorithms (all of which unless otherwise stated are described in \citet{kaufman2009finding}) were used, all with standard R-implementations and default settings:

\begin{itemize}
	\item Partitioning Around Medoids (PAM),
	\item single linkage,
	\item average linkage,
	\item complete linkage,
	\item Ward's method (this was originally defined for Euclidean data but can be generalised to general dissimilarities, see \citet{MurLeg14}),
	\item spectral clustering (\cite{NgJoWe01}).
\end{itemize}

\section{A composite cluster validity index based on indexes measuring different aspects of clustering}
\label{sec:aggregation_indexes}

\subsection{Cluster validity indexes}
\label{sec:indexes}

In order to choose a clustering method and number of clusters for clustering the players, we will follow the concept of aggregation of calibrated cluster validity indexes as introduced in \cite{hennig2017cluster} and elaborated in \cite{akhanli2020comparing}.

A large number of cluster validity indexes are proposed in the literature, for example the Average Silhouette Width (ASW) \citep{kaufman2009finding}, the Calinski-Harabasz index (CH) \citep{calinski1974dendrite}, the Dunn index \citep{dunn1974well}, a Clustering Validity Index Based on Nearest Neighbours (CVNN) \citep{liu2013understanding}, and Hubert's $\Gamma$ \citep{hubert1976quadratic}. All these indexes attempt to summarise the quality of a clustering as a single number. They are normally optimised in order to find the best clustering out of several clusterings. Mostly the set of compared clusterings is computed from the same clustering method but with different numbers of clusters. Clusterings computed by different methods can also be compared in this way, but this is done much less often, and some indexes are closer connected to specific clustering methods than others (e.g., optimising CH for a fixed number of clusters is equivalent to $k$-means). See \cite{AGMPP12} for a comparative simulation study, and \cite{HVH15} for more indexes and discussion. The indexes are usually presented as attempts to solve the problem of finding the uniquely best clustering on a data set. Occasionally the ASW is also used to assess a clustering's validity without systematic optimisation. Alternatively, stability under resampling has been suggested as a criterion for measuring the quality of a clustering (\cite{tibshirani2005cluster, fang2012selection}). Further approaches to choose the number of clusters are more closely related to specific clustering methods and their objective functions, such as the gap statistic \citep{TiWaHa01}. In model-based clustering, information criteria such as the BIC are popular \citep{BCMR19}. As the indexes above, these are also usually interpreted as stand-alone measures of the clustering quality. 

As argued in \cite{hennig2015clustering31,hennig2015true}, there are various aspects of clusterings that can be of interest, such as separation between clusters, within-cluster homogeneity in the sense of small within-cluster dissimilarities or homogeneous distributional shapes, representation of clusters by their centroids, stability under resampling, and entropy. In many situations two or more of these aspects are in conflict; for example single linkage clustering will emphasise between-cluster separation disregarding within-cluster homogeneity, whereas complete linkage will try to keep within-cluster dissimilarities uniformly small disregarding separation. In different applications, different aspects of clustering are of main interest, and there can be different legitimate clusterings on the same data set depending on which characteristics are required. For example, different biological species need to be genetically separated, whereas within-cluster homogeneity is often more important than separation for example when colouring a map for highlighting clusters of similar regions according to criteria such as economic growth, severity of a pandemic, or avalanche risk.

The chosen clustering then needs to depend on a user specification of relevant features of the clustering. The traditional literature on validity indexes gives little guidance in this respect; where such indexes are introduced, authors tend to argue that their new index is the best over a wide range of situations, and comparative studies such as \cite{AGMPP12} normally focus on the ability of the indexes to recover a given ``true'' clustering. The approach taken here is different. It is based on defining indexes that separately measure different aspects of clustering quality that might be of interest, and the user can then aggregate the indexes, potentially involving weights, in order to find a clustering that fulfills the specific requirements of a given application.

In the following we will first define indexes that measure various characteristics of a clustering that are potentially of interest for the clustering of football players, and then we will propose how they can be aggregated in order to define an overall index that can be used to assess clusterings and select an optimal one. 

\subsection{Measurement of individual aspects of clustering quality}
\label{subsubsec:aspect_cquality}

\cite{hennig2017cluster} and \cite{akhanli2020comparing} defined several indexes that measure desirable characteristics of a clustering (and contain more details than given below). Not all of these are relevant for clustering football players. We will define the indexes that are later used in the present work, and then give reasons why further indexes have not been involved. 

\begin{description}
	\item[Average within-cluster dissimilarities:] This index formalises within-cluster homogeneity in the sense that observations in the same cluster should all be similar. This is an essential requirement for useful clusters of football players.
	
\begin{displaymath}
	I_{ave.within}(\mathcal{C}) = \frac{1}{n} \sum_{k=1}^{K} \frac{1}{n_k-1}\sum_{x_{i} \neq x_{j} \in C_{k}} d(x_{i},x_{j}).
\end{displaymath}
	
A smaller value indicates better clustering quality. 

\item[Separation index:] Objects in different clusters should be different from each other. This is to some extent guaranteed if the within-cluster dissimilarities are low (as then the larger dissimilarities tend to be between clusters), but usually, on top of this, separation is desirable, meaning that there is some kind of gap between the clusters. The idea is that clusters should not just result from arbitrarily partitioning a uniformly or otherwise homogeneously distributed set of observations. There is no guarantee that there is meaningful separation between clusters in the set of football players, but if such separation exists between subsets, these are good cluster candidates. Separation refers to dissimilarities between observations that are at the border of clusters, and closer to other clusters than the interior points of clusters. Therefore, separation measurement is based on the observations that have smallest dissimilarities to points in other clusters.
 	
For every object $x_{i} \in C_{k}$, $i = 1, \ldots, n$, $k \in {1, \ldots, K}$, let $d_{k:i} = \min_{x_{j} \notin C_{k}} d(x_{i},x_{j})$. Let $d_{k:(1)} \leq \ldots \leq d_{k:(n_{k})}$ be the values of $d_{k:i}$ for $x_{i} \in C_{k}$ ordered from the smallest to the largest, and let $[pn_{k}]$ be the largest integer $\leq pn_{k}$. Then, the separation index with the parameter $p$ is defined as
	
	\begin{displaymath}
		I_{sep}(\mathcal{C};p) = \frac{1}{\sum_{k=1}^{K} [pn_{k}]} \sum_{k=1}^{K} \sum_{i=1}^{[pn_{k}]} d_{k:(i)},  
	\end{displaymath}
	
Larger values are better. The proportion $p$ is a tuning parameter specifying what percentage of points should contribute to the ``cluster border''. We suggest $p=0.1$ as default.  

\item[Representation of dissimilarity structure by the clustering:] A clustering can be seen as a parsimonious representation of the overall dissimilarities. In fact, a clustering of football players can be used as a simplification of the dissimilarity structure by focusing on players in the same cluster rather than using the exact dissimilarities to consider more or less similar players. The quality of a clustering as representation of the dissimilarity structure can be measured by several versions of the family of indexes known as Hubert's $\Gamma$ introduced by  \cite{hubert1976quadratic}. The version that can be most easily computed for a data set of the given size is based on the Pearson sample correlation $\rho$. It interprets the ``clustering induced dissimilarity'' $\mathbf{c} = vec([c_{ij}]_{i<j})$, where $c_{ij} = \mathbf{1}(l_{i} \neq l_{j})$, i.e. the indicator whether $x_i$ and $x_j$ are in different clusters, as a ``fit'' of the given data dissimilarity $\mathbf{d} = vec\left([d(x_{i}, x_{j})]_{i<j}\right)$, and measures its quality as

	\begin{displaymath}
		I_{Pearson \Gamma}(\mathcal{C}) = \rho(\mathbf{d}, \mathbf{c}).
	\end{displaymath}	
	
This index has been used on its own to measure clustering quality, but we use it as measuring a specific aspect of clustering quality. Large values are good.

\item[Entropy:] Although not normally seen as primary aim of clustering, in some applications very small clusters are not very useful, and cluster sizes should optimally be close to uniform. This is measured by the well known ``entropy'' \cite{shannon1948mathematical}: 

	\begin{displaymath}
		I_{entropy}(\mathcal{C}) = - \sum_{k=1}^{K} \frac{n_{k}}{n} \log(\frac{n_{k}}{n}).
	\end{displaymath}
	
Large values are good. For the clustering of football players, we aim at a high entropy, as too large clusters will not differentiate sufficiently between players, and very small clusters (with just one or two players, say) are hardly informative for the overall structure of the data.  

\item[Stability:] Clusterings are often interpreted as meaningful if they can be generalised as stable substantive patterns. Stability means that they can be replicated on different data sets of the same kind. Without requiring that new independent data are available, this can be assessed by resampling methods such as cross-validation and bootstrap. 

It is probably not of much interest to interpret the given set of football players as a random sample representing some underlying true substantially meaningful clusters that would also be reproduced by different players. However, it is relevant to study the stability of the clustering of football players under resampling, as such stability means that whether certain players tend to be clustered together does not depend strongly on which other players are in the sample, which is essential for interpreting the clusters as meaningful. 

Two approaches from the literature have been used for clustering stability measurement in  \cite{akhanli2020comparing}, namely the prediction strength \cite{tibshirani2005cluster}, and a bootstrap-based method (called ``Bootstab'' here) by
\citet{fang2012selection}. We focus on the latter below. In the original paper this (as well as the prediction strength) was proposed for assessing clustering quality and making decisions such as regarding the number of clusters on their own, but this is problematic. Whereas it makes sense to require a good clustering to be stable, it cannot be ruled out that an undesirable clustering is also stable. We therefore involve Bootstab as measuring just one of several desirable clustering characteristics. 

  
$B$ times two bootstrap samples are drawn from the data with replacement. Let $X_{[1]},\ X_{[2]}$ the two bootstrap samples in the $b$th bootstrap iteration. For $t=1, 2,$ let $L_{b}^{(t)} = \left( l_{1b}^{(t)}, \ldots, l_{nb}^{(t)} \right)$ based on the clustering of $X_{[t]}$. This means that for points $x_i$ that are resampled as member of $X_{[t]}$, $l_{ib}^{(t)}$ is just the cluster membership indicator, whereas for points $x_i$ not resampled as member of $X_{[t]}$, $l_{ib}^{(t)}$ indicates the cluster on $X_{[t]}$ to which $x_i$ is classified using a suitable supervised classification method (we use the methods listed in \cite{akhanli2020comparing}, extending the original proposal in \cite{fang2012selection}). The Bootstab index is

\begin{displaymath}
	I_{Bootstab}(\mathcal{C}) = \frac{1}{B} \sum_{b=1}^{B} \left\{ \frac{1}{n^2} \sum_{i,i'} \left|f_{ii^{'}b}^{(1)} - f_{ii^{'}b}^{(2)}\right| \right\},
\end{displaymath}

\noindent where for $t=1,2$,

\begin{displaymath}
	f_{ii^{'}b}^{(t)} = \mathbf{1} \left( l_{i'b}^{(t)}= l_{ib}^{(t)} \right),
\end{displaymath}

\noindent indicating whether $x_i$ and $x_{i'}$ are in or classified to the same cluster based on the clustering of $X_{[1t]}$. $I_{Bootstab}$ is a percentage of pairs that have different ``co-membership'' status based on clusterings on two bootstrap samples. Small values of $I_{Bootstab}$ are better. 
\end{description}

The following indexes from \cite{hennig2017cluster} are not involved here, because they seem rather irrelevant to potential uses of clusters of football players: representation of clusters by centroids; small within-cluster gaps; clusters corresponding to density modes; uniform or normal distributional shape of clusters.   

\subsection{Aggregation and calibration of indexes}
\label{subsec:aggregation_indexes}

Following \cite{akhanli2020comparing}, indexes measuring different desirable aspects of a clustering are aggregated computing a weighted mean. For selected indexes $I^*_{1}, \ldots, I^*_{s}$ with weights $w_{1}, \ldots, w_{s} > 0$:

\begin{equation}
	\mathcal{A}(\mathcal{C}) = \frac{\sum_{j=1}^{s} w_{j} I^*_{j}(\mathcal{C})}{\sum_{j=1}^{s} w_{j}}.
	\label{eq:aggregation_indexes}
\end{equation} 

The weights are used to up- or down-weight indexes that are more or less important than the others for the aim of clustering in the situation at hand. This assumes that all involved indexes are calibrated so that their values are comparable and that they point in the same direction, e.g., that large values are better for all of them. The latter can be achieved easily by multiplying those indexes that are better for smaller values by $-1$.  

The following approach is used to make the values of the different indexes comparable. We generate a large number $m$ 
of random clusterings $\mathcal{C}_{R1},\ldots,\mathcal{C}_{Rm}$ on the data. On top of these there are $q$ clusterings produced by regular clustering methods as listed in Section \ref{sec:cmethods}, denoted by ${\mathcal C}_1,\ldots,\mathcal{C}_q$. For given data set $\mathcal{X}$ and index $I$, the clusterings are used to standardise $I$:

\begin{eqnarray*}
	m(I,\mathcal{X})&=&\frac{1}{m+q}\left(\sum_{i=1}^m I(\mathcal{C}_{Ri})+ \sum_{i=1}^q I(\mathcal{C}_{i})\right),\\ 
	s^2(I,\mathcal{X})&=& \frac{1}{m+q-1}\left(\sum_{i=1}^m \left[I(\mathcal{C}_{Ri})-
	m(I,\mathcal{X})\right]^2+ \sum_{i=1}^q \left[I(\mathcal{C}_{i})-m(I,\mathcal{X})\right]^2\right),\\
	I^*(\mathcal{C}_{i})&=&\frac{I(\mathcal{C}_i)-m(I,\mathcal{X})}{s(I,\mathcal{X})},\ 
	i=1,\ldots,q.
\end{eqnarray*}

$I^*$ is therefore scaled so that its values can be interpreted as expressing the quality compared to what the collection of clusterings $\mathcal{C}_{R1},\ldots,\mathcal{C}_{Rm},{\mathcal C}_1,\ldots,\mathcal{C}_q$ achieves on the same data set. The approach depends on the definition of the random clusterings. These should generate enough random variation in order to work as a tool for calibration, but they also need to be reasonable  as clusterings, because if all random clusterings are several standard deviations away from the clusterings provided by the standard clustering methods, the  exact distance may not be very meaningful.

Four different algorithms are used for generating the random clusterings, ``random $K$-centroids'', ``random nearest neighbour'', ``random farthest neighbour'', and ``random average distances'', for details see \cite{akhanli2020comparing}.

Assume that we are interested in numbers of clusters $K\in\{2,\ldots,K_{max}\}$, and that all clustering methods of interest are applied for all these numbers of clusters. Section \ref{sec:cmethods} lists six clustering methods, and there are four approaches to generate random clusterings. Therefore we compare $q=6(K_{max}-1)$ clusterings from the methods and  $m=4B(K_{max}-1)$ random clusterings, where $B=100$ is the number of random clusterings generated by each approach for each $K$. 

Two different ways to calibrate the index values have been proposed in \cite{akhanli2020comparing}:
\begin{description}
	\item[C1:] All index values can be calibrated involving clusterings with all numbers of clusters.
	\item[C2:] Index values for a given number of clusters $k$ can be calibrated involving only clusterings with $k$ clusters. 
\end{description}
In order to understand the implications of these possibilities it is important to note that some of the indexes defined in Section \ref{subsubsec:aspect_cquality} will systematically favour either larger or smaller numbers of clusters. For example, a large number of clusters will make it easier for $I_{ave.within}$ to achieve better values, whereas a smaller number of clusters will make it easier  for $I_{sep}$ to achieve better values. Option C1 will not correct potential biases of the collection of involved indexes in favour of larger or smaller numbers of clusters. It is the method of choice if any tendency in favour of larger or smaller numbers of clusters implied by the involved indexes is desired, which is the case if the indexes have been chosen to reflect desirable characteristics of the clusterings regardless of the number of clusters. Option C2 employs the involved indexes relative to the number of clusters, and will favour a clustering that stands out on its specific number of clusters, even if not in absolute terms. When using option C1, the choice of the number of clusters is more directly determined by the chosen indexes, whereas calibration according to option C2 will remove systematic tendencies of the indexes when choosing the number of clusters, and can therefore be seen as a more data driven choice. 

\section{Application to the football player data}
\label{sec:valresults}

The clustering methods listed in Section \ref{sec:cmethods} will be applied to the football player data set using a range of numbers of clusters. The quality of the resulting clusterings is measured and compared according to the composite cluster validity index $\mathcal{A}$ as defined in (\ref{eq:aggregation_indexes}). The involved indexes are $I^*_1=I^*_{ave.within}, I^*_2=I^*_{sep}, I^*_3=I^*_{Pearson \Gamma}, I^*_4=I^*_{entropy}, I^*_5=I^*_{Bootstab}$, see Section \ref{subsubsec:aspect_cquality}, where the upper star index means that indexes are calibrated, see Section \ref{subsec:aggregation_indexes}. 

Corresponding to the two different aims of clustering as outlined in Section \ref{sec:intro}, two different sets of weights $w_1,\ldots,w_5$ will be used. 

\subsection{A data driven composite index}
\begin{figure}[tbp]
  \centering
  \includegraphics[width=0.48\textwidth]{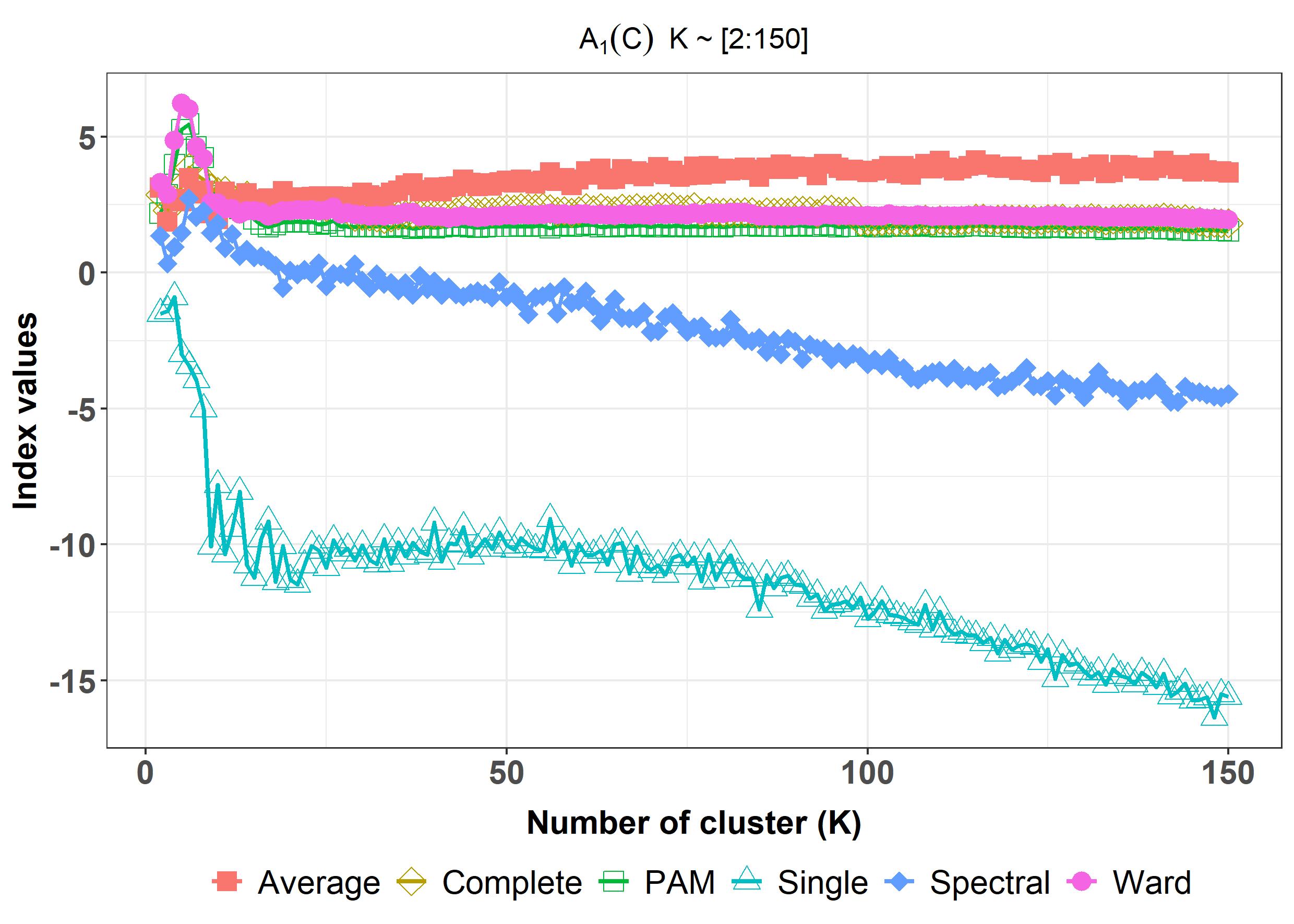}
  \includegraphics[width=0.48\textwidth]{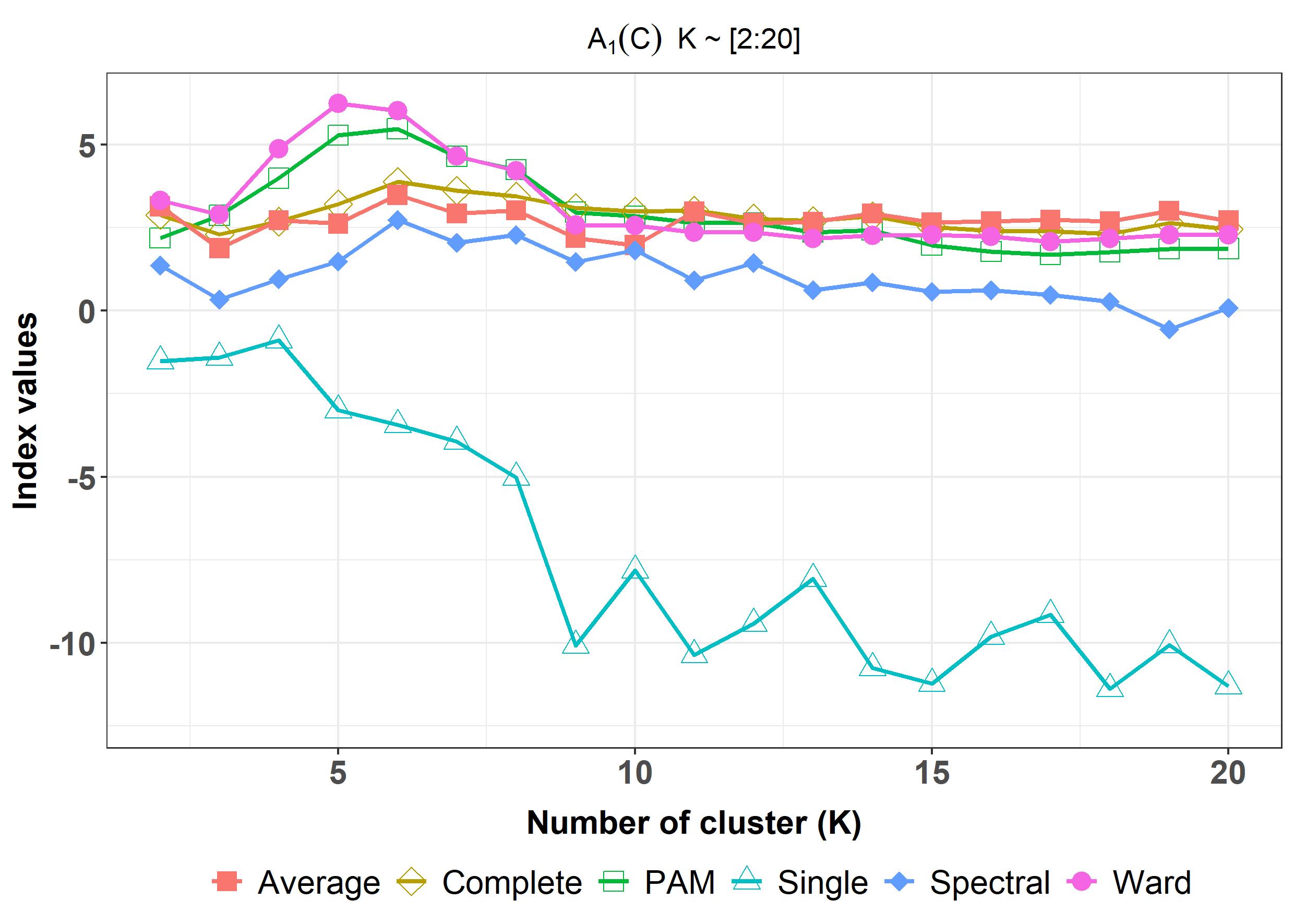}
  	\caption{Results for football data with calibration index $\mathcal{A}_{1}(C)=I_{ave.wit} + 0.5I_{sep.index} + I_{Pearson \Gamma} + I_{entropy} + I_{Bootstab}$. Left side: full range of the number of clusters; right side: number of clusters in the range $[2:20]$.}
  	\label{fig:a1}	
\end{figure}

The first clustering is computed with the aim of giving a raw representation of inherent grouping structure in the data. For this aim we choose calibration strategy C2 from Section \ref{subsec:aggregation_indexes}. A first intuitive choice of weights, given that the five involved indexes all formalise different desirable features of the clustering, would be $w_1=w_2=w_3=w_4=w_5=1$ (W1). Experience with the working of the indexes suggests that $I^*_{sep}$ has a tendency to favour clusterings that isolate small groups or even one point clusters of observations. It even tends to yield better values if the remainder of the observations is left together (as splitting them up will produce weaker separated clusters). Although a certain amount of separation is desirable, it is advisable to downweight $I^*_{sep}$, as it would otherwise go too strongly against the requirements of small within-cluster distances and entropy, which are more important. Similarity of the players in the same cluster is a more elementary feature for interpreting the clusters, and the clustering should differentiate players properly, which would not be the case if their sizes are too imbalanced. For this reason we settle for $\mathcal{A}_{1}(\mathcal{C})$ defined by $w_2=\frac{1}{2},\  w_1=w_3=w_4=w_5=1$ (W2). The optimal clustering, the five cluster solution of Ward's method, is in fact the same for W1 and W2, but the next best clusterings are different, and the best clusterings stick out quite clearly using $\mathcal{A}_{1}(\mathcal{C})$, see Figure \ref{fig:a1} and Table \ref{tab:football_data_validitiy_index_comparison} (note that the listed values of $\mathcal{A}_{1}(\mathcal{C})$ and $\mathcal{A}_{2}(\mathcal{C})$ as defined below can be interpreted in terms of the standard deviations per involved index compared to the set of clusterings used for calibration).

\begin{figure}[htbp]
  \centering
  \includegraphics[width=1\textwidth]{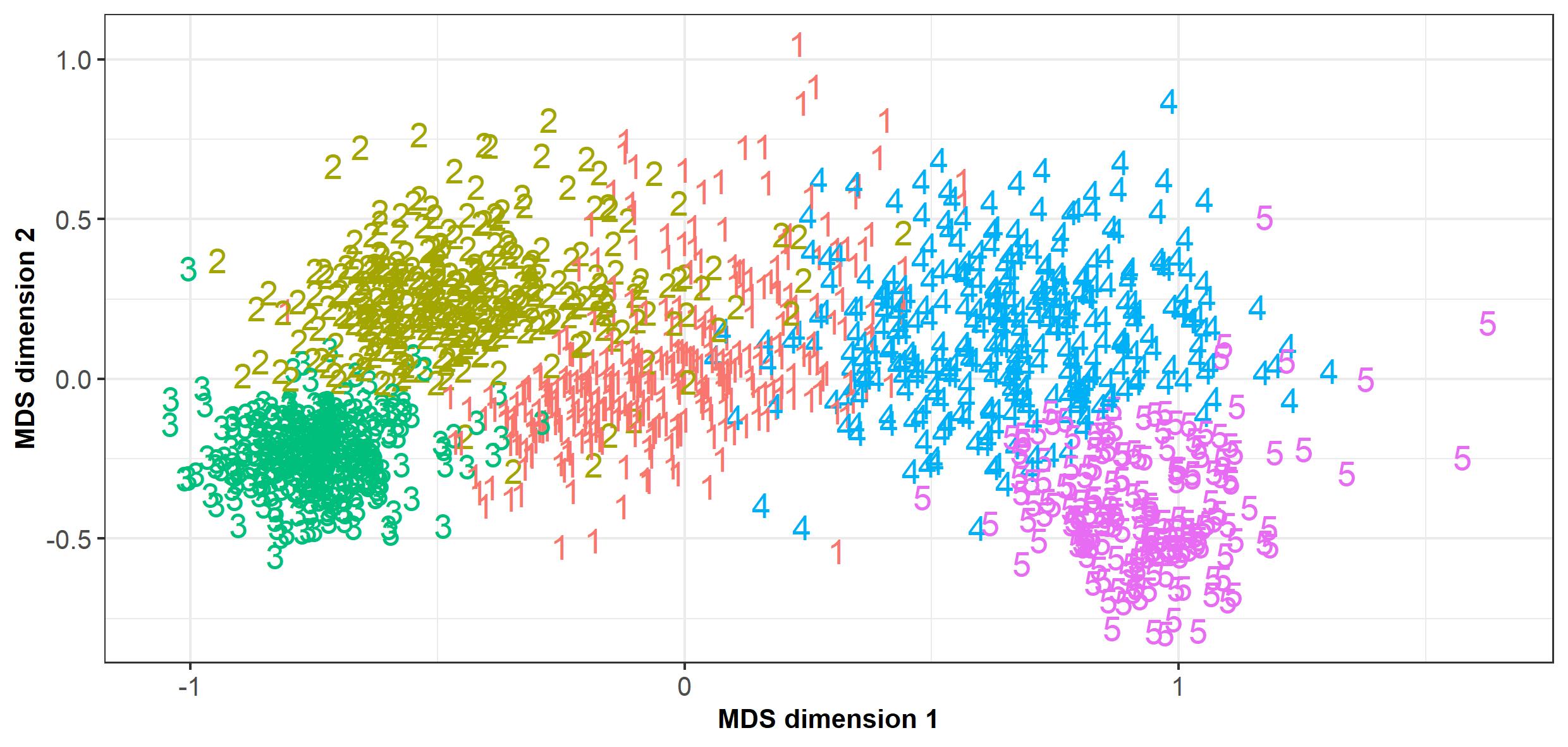}
 \caption{Multidimensional scaling representation of the data with Ward clustering, $K=5$.}
  \label{fig:ward_mds_pam_51}
\end{figure}

\begin{figure}[htbp]
  \centering
   \includegraphics[width=1\textwidth]{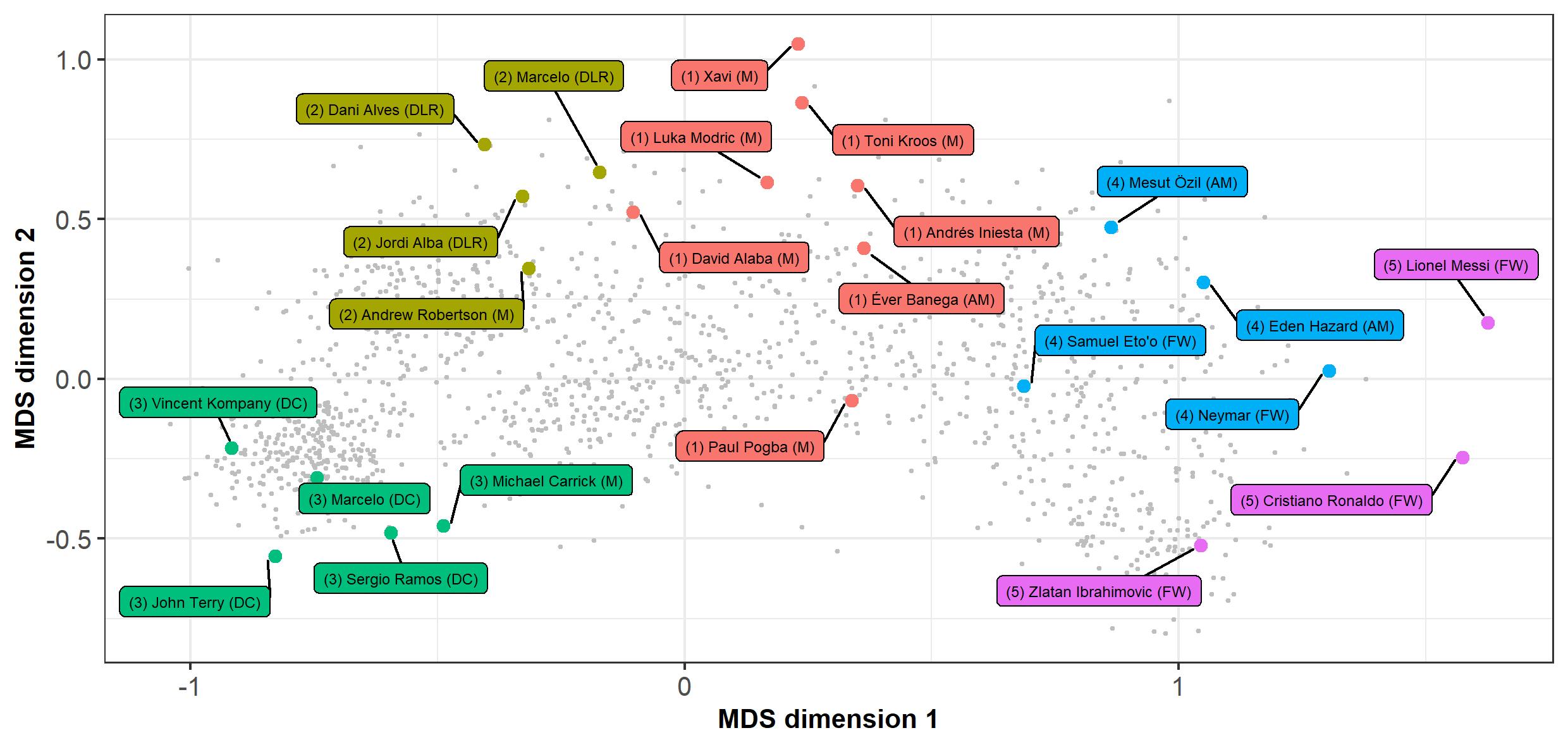}
\caption{Multidimensional scaling representation of the data with Ward clustering, $K=5$, with location of some well known players.}
  \label{fig:ward_mds_pam_52}
\end{figure}

A visualisation of the clustering using MDS is in Figures~\ref{fig:ward_mds_pam_51}, \ref{fig:ward_mds_pam_52}. Commenting on clusters from left to right in the MDS plot, corresponding to going from defensive to offensive players, cluster 3 mainly contains centre backs (DC), cluster 2 mainly contains full backs (DR or DL), cluster 1 mainly involves midfielders (M), cluster 4 has attacking midfielders (AM), and cluster 5 mainly contains forwards (FW), respectively. Table~\ref{tab:top_level_summary} in the Appendix gives cluster-wise comprehensive statistical summaries for the top level performance variables. Cluster 3 is characterised by strong values in defensive features, such as interceptions, clearances, aerial duels and long passes. Somewhat surprisingly they also do most free kicks. Cluster 2 players are on average strongest in blocks, and good at cross passes compared with the other more defensive clusters. They are weakest at scoring goals. Players in cluster 1 are on average the strongest in tackles and short passes. Otherwise their values are in between the two more defensive and the two more offensive clusters 4 and 5. Players in cluster 4 support the goalscorers, who mainly are in cluster 5. In cluster 4, players have most dribbles, crosses, key passes, assists, fouls given in their favour, tend to play most corners, but are also dispossessed most. Cluster 5 leads regarding shots and goals, but these players also commit most fouls, are most often in an offside position, have most unsuccessful touches, and have the clearly lowest values regarding passes.   

The clusters are strongly aligned with the players’ positions, but they are not totally dominated by these positions. For instance, cluster 1 mainly contains defensive midfielders, but some players are in different positions, such as Banega. Although he is usually deployed as a central midfielder, he is well capable to play as an attacking one. Banega was engaged as defensive midfielder in Boca Juniors, but his technical skills, such as dribbling ability, quick feet, vision and accurate passing enabled him to play as a attacking midfielder \citep{everbanega1}. His historical background and his playing style placed him in cluster 1. Another example is Carrick, who is a midfielder, but his style of play relies on defensive roles, such as tackles, stamina, physical attributes, etc. \citep{michealcarrick}. These kinds of playing characteristics put him into cluster 3, which mainly contains central defenders. Samuel Eto’o is a forward player and could as such be expected in cluster 5, but his playing style rather fits in cluster 4, which mostly includes attacking midfielders. During Inter's 2009–10 treble-winning season, Eto'o played an important role in the squad, and was utilised as a winger or even as an attacking midfielder on the left flank in Mourinho's 4–2–3–1 formation, where he was primarily required to help his team creatively and defensively with his link-up play and work-rate off the ball, which frequently saw him tracking back \citep{samueletoo}. 

\subsection{A composite index for smaller clusters based on expert assessments}
The second clustering is computed with the aim of having smaller homogeneous clusters that unite players with very similar characteristics. These can be used by managers for finding players that have a very similar profile to a given player, and for characterising the team composition at a finer scale. Larger numbers of clusters become computationally cumbersome for assessing stability and for the resampling scheme introduced in Section \ref{subsec:aggregation_indexes}. For this reason the maximum investigated number of clusters is 150; we assume that clusters with 10 players on average deliver a fine enough partition. In fact very small clusters with, say, 1-3 players, may not be very useful for the given aim, or only for very exceptional players. 

In order to find a suitable weighting for a composite index we conducted a survey of 13 football experts. The idea of the survey was to have several questions, in which alternatives are offered to group a small set of famous players. The experts were then asked to rank these groupings according to plausibility. The groupings were chosen in order to distinguish between different candidate clusterings from the methods listed in Section \ref{sec:cmethods} between 100 and 150 clusters (single linkage and spectral clustering were not involved due to obvious unsuitability, in line with their low value on the resulting composite index). 

More precisely, different clustering solutions correspond to the multiple choices in each question, and each selection is based on a different clustering solution. For the selected players for the survey, these groupings do not change over ranges of numbers of clusters; e.g., for PAM with $K \in \{100,\ldots,113\}$, see Table~\ref{tab:survey_res_cluster_selections}. The respondents answer each question by ranking different groupings in order of plausibility from 1 to the number of multiple choices of that question. The questions are presented in the Appendix.

We have collaborated with The İstanbul Başakşehir football club. The survey questions were asked to 13 football experts including the head coach, the assistant coaches, the football analysts and the scouts of this club, and some journalists who are experienced with European football. 

For the ranking responses of the survey questions we assigned scores for each rank in each question, where the score assignment was made in a balanced way, because each question has a different number of possible choices. Table~\ref{tab:point_assignment} shows the assignment of the scores. The idea behind the scoring system is that a question with five choices gives more differentiated information; the score difference between the first rank and the last rank is therefore bigger than for questions with fewer choices, however the difference between first and second rank should be bigger for a lower number of choices, as with five choices the quality of the best two is more likely assessed as similar, as both of these are ranked ahead of further choices, whereas with two choices overall this is not the case. Apart from these considerations, as we were interested in the comparison between all choices by the experts rather than focusing on their favourites, score differences between adjacent ranks have been chosen as constant given the same number of choices in the question.    

\begin{table}[htbp]
	\caption{Clustering selections with the clustering algorithms and their number of clusters range }
	\begin{tabular}{l | ll}
	\thickhline 
	\textbf{Selections}     & \textbf{Clustering Algorithms} & \textbf{Number of clusters range}    \\
    \hline	
	Selection 1 & PAM                            & $K \sim [100:113]$                   \\
	Selection 2 & PAM                            & $K \sim [114:118]$                   \\
	Selection 3 & PAM                            & $K \sim [119:129, 134:136, 147:150]$ \\
	Selection 4 & PAM                            & $K \sim [130:133, 137:146]$          \\
	Selection 5 & Ward's method                  & $K \sim [100:147]$                   \\
	Selection 6 & Ward's method                  & $K \sim [148:150]$                   \\
	Selection 7 & Complete linkage               & $K \sim [100:150]$                   \\
	Selection 8 & Average linkage                & $K \sim [100:150]$                   \\
	\thickhline  	
	\end{tabular}
	\label{tab:survey_res_cluster_selections}	
\end{table}

\begin{table}[htbp]
	\caption{Score assignment for the survey questions}
	\begin{tabular}{c | c c c c c}
	\thickhline 
	\textbf{The selection of multiple choices} & \textbf{1. Rank} & \textbf{2. Rank} & \textbf{3. Rank} & \textbf{4. Rank} & \textbf{5. Rank} \\
        \hline
	5 choices & 30 & 24 & 18 & 12 & 6 \\
	3 choices & 30 & 20 & 10 & -  & - \\
	2 choices & 30 & 15 & -  & -  & - \\
	\thickhline 
	\end{tabular}
	\label{tab:point_assignment}	
\end{table}

\begin{table}[tbp]
	\caption{Total scores of the seven survey questions for different clustering selections from each of the 13 football experts.}
	\begin{tabular}{l | cccccccc}
	\thickhline 
	
	\multirow{2}{*}{Respondents} & \multicolumn{8}{c}{\underline{Selection}} \\
	& 1 & 2 & 3 & 4 & 5 & 6 & 7 & 8 \\
    \hline
	Head coach        			 & 138 & 138 & 162 & 162 & 148 & 160 & 109 & 125 \\ 
	Assistant coach - 1		     & 138 & 138 & 144 & 144 & 144 & 166 & 109 & 137 \\ 
	Assistant coach - 2 		 & 125 & 115 & 127 & 137 & 109 & 121 & 136 & 134 \\ 
	Goalkeeping coach 			 & 148 & 118 & 130 & 160 & 152 & 176 & 109 & 125 \\ 
	Individual performance coach & 166 & 136 & 148 & 178 & 146 & 152 & 109 & 119 \\
	Physical performance coach   & 159 & 149 & 119 & 129 & 125 & 137 & 116 & 168 \\
	Football Analyst			 & 132 & 132 & 144 & 144 & 166 & 154 & 123 & 139 \\
	Chief Scout	                 & 176 & 166 & 166 & 176 & 134 & 128 & 117 & 155 \\
	Scout - 1                    & 144 & 144 & 150 & 150 & 154 & 148 & 99  & 97  \\
	Scout - 2                    & 113 & 143 & 155 & 125 & 133 & 145 & 142 & 168 \\
	Scout - 3                    & 148 & 118 & 100 & 130 & 132 & 126 & 115 & 129 \\
	Journalist - 1               & 129 & 149 & 161 & 141 & 95  & 123 & 150 & 156 \\
	Journalist - 2               & 154 & 134 & 116 & 166 & 136 & 160 & 117 & 145 \\
	\hline
	TOTAL                        & 1870 & 1780 & 1822 & 1942 & 1774 & 1896 & 1531 & 1797 \\
	\thickhline  	
	\end{tabular}
	\label{tab:survey_res}	
\end{table}

Table~\ref{tab:survey_res} shows the result of the survey based on the responses from each expert. It shows substantial variation between the experts. As a validation, we conducted a test of the null hypothesis $H_0$ of randomness of the experts' assessments. The $H_0$ was that the experts assigned ranks to the alternative choices randomly and independently of each other. The test statistic was the resulting variance of the sum scores of the eight selections listed in Table \ref{tab:survey_res_cluster_selections}.  In case that there is some agreement among the experts about better and worse selections, the variance of the sum scores should be large, as higher ratings will concentrate on the selections agreed as better, and lower ratings will concentrate on the selections agreed as worse. The test is therefore one-sided. The distribution of the test statistic under $H_0$ was approximated by a Monte Carlo simulation of 2000 data sets
\citep{Marriott79}, in which for each expert random rankings for all the survey questions were drawn independently. This yielded $p=0.048$, just about significant at the 5\% level. Although not particularly convincing, this at least indicates some agreement between the experts.  

According to the survey, the clusterings of Selection 4 are best, but due to the considerable disagreement between the experts and the limited coverage of the overall clusterings by the survey questions, we use the survey result in a different way rather than just taking Selection 4 as optimal. Instead, we choose a weighting for a composite index $\mathcal{A}_{2}$ that optimises the Spearman correlation between the value of $\mathcal{A}_{2}(\mathcal{C})$, for each selection maximised over the clusterings in that selection, and the selection's sum scores from the survey as listed in the last line of Table  \ref{tab:survey_res}. We believe that the resulting composite index represents the experts' assessments better than just picking a clustering from Selection 4, particularly if applied to future data of the same kind, because it allows to generalise the assessments beyond the limited set of players used in the survey questions. 

Although we did not run a formal optimisation, the best value of 0.524 that we found experimentally was achieved for $w_1=w_2=w_3=0,\ w_4=0.5,\ w_5=1$. $I^*_{Bootstab}$ is the only index to  favour PAM solutions with large $K$, and these are ranked generally highly by the sum scores, so it is clear that $w_5$, the weight for  $I^*_{Bootstab}$, must be high. In fact, using $I^*_{Bootstab}$ alone achieves the same Spearman correlation value of 0.524, but if $I^*_{Bootstab}$ is used on its own, useless single linkage solutions with 2 and 3 clusters are rated as better than the best PAM solutions with $K>100$, whereas the composite index with $w_4=0.5$ makes the latter optimal over the whole range of $K$. Spearman rather than Pearson correlation was used, because the Pearson correlation is dominated too strongly by the outlyingly bad rating for Selection 7. The majority of indexes, including all indexes proposed in the literature for stand-alone use presented in Table~\ref{tab:football_data_validitiy_index_comparison} (which includes the best results found by $\mathcal{A}_{2}(\mathcal{C})$), yield negative Spearman correlations with the expert's sum scores; entropy on its own achieves a value of 0.214.

\begin{figure}[tbp]
  \centering
  \includegraphics[width=0.48\textwidth]{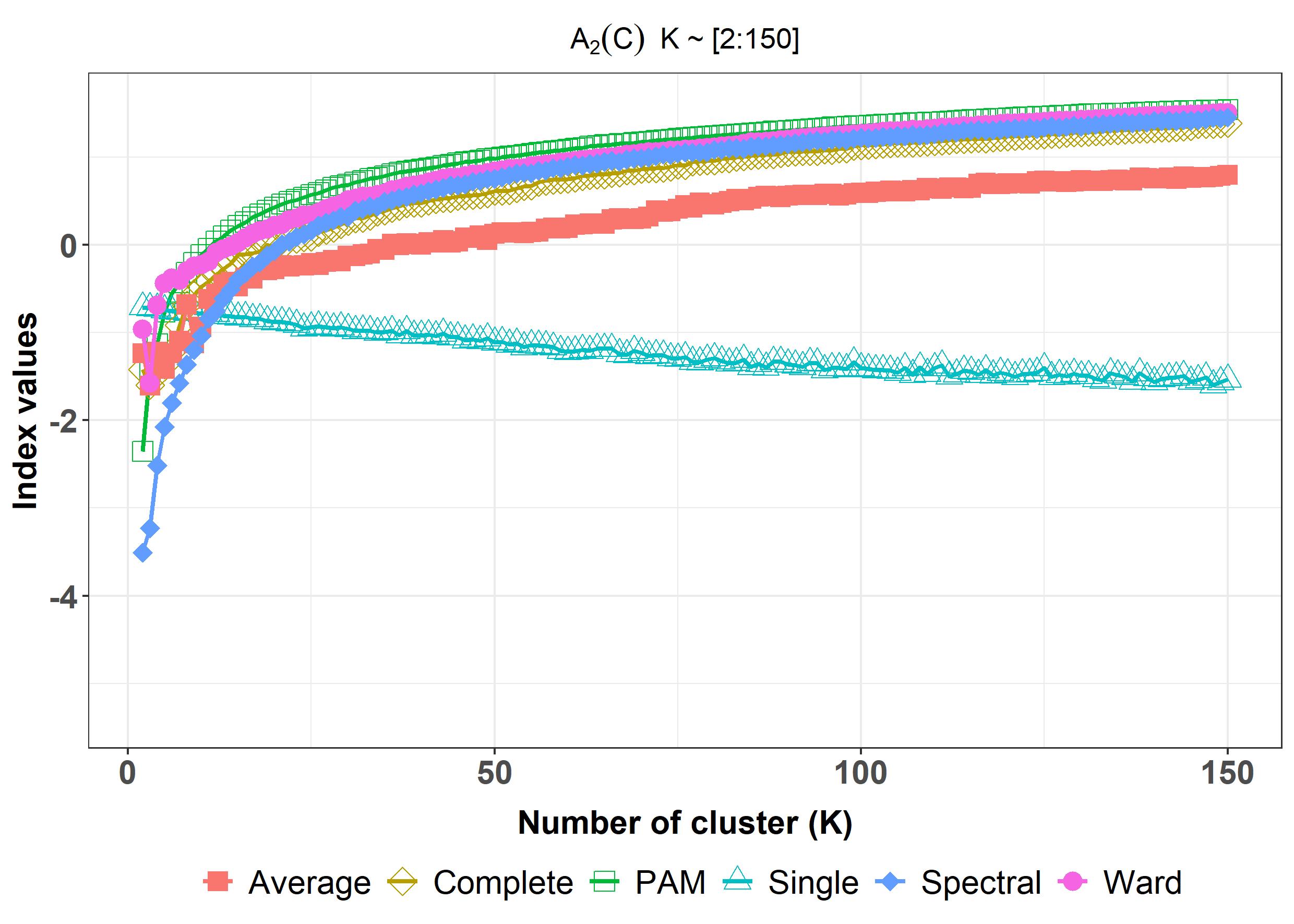}
  \includegraphics[width=0.48\textwidth]{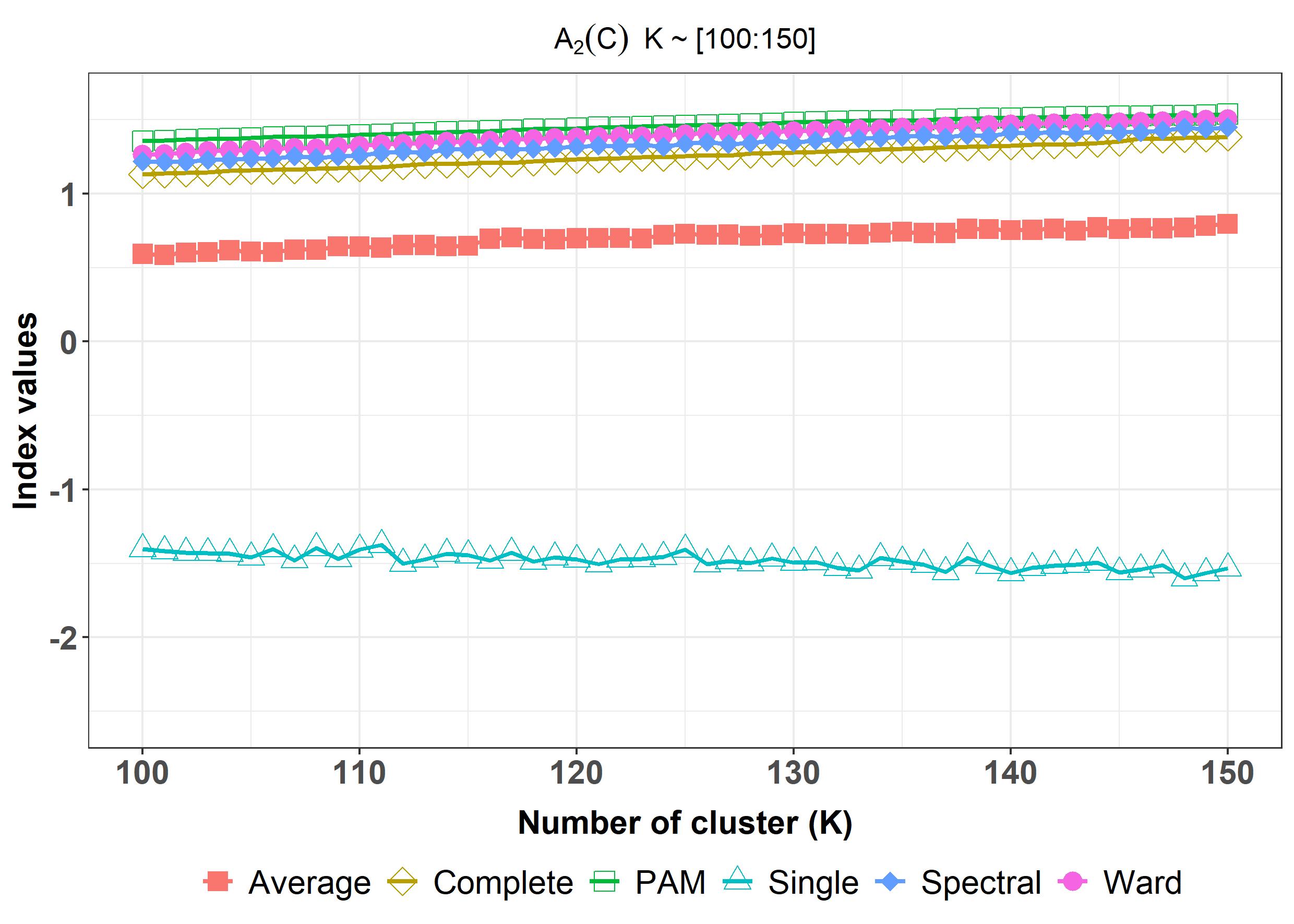}
  	\caption{Results for football data with the calibration index $\mathcal{A}_2(C)=0.5I_{entropy} + I_{Bootstab}$. Left side: full range of the number of clusters; right side: number of clusters in the range $[100:150]$.}
  	\label{fig:a2}	
\end{figure}  

According to $\mathcal{A}_{2}(\mathcal{C})$ with weights as above, the best clustering is PAM with $K=150$ from Selection 3. This has an ARI of 0.924 when compared with the PAM solution with $K=146$, which belongs to Selection 4, optimal according to the expert's sum score, so these clusterings are very similar (this is the highest ARI value among the ARIs between the best two clusterings of any two Selections). 

\begin{figure}[tbp]
  \centering
  \includegraphics[width=1\textwidth]{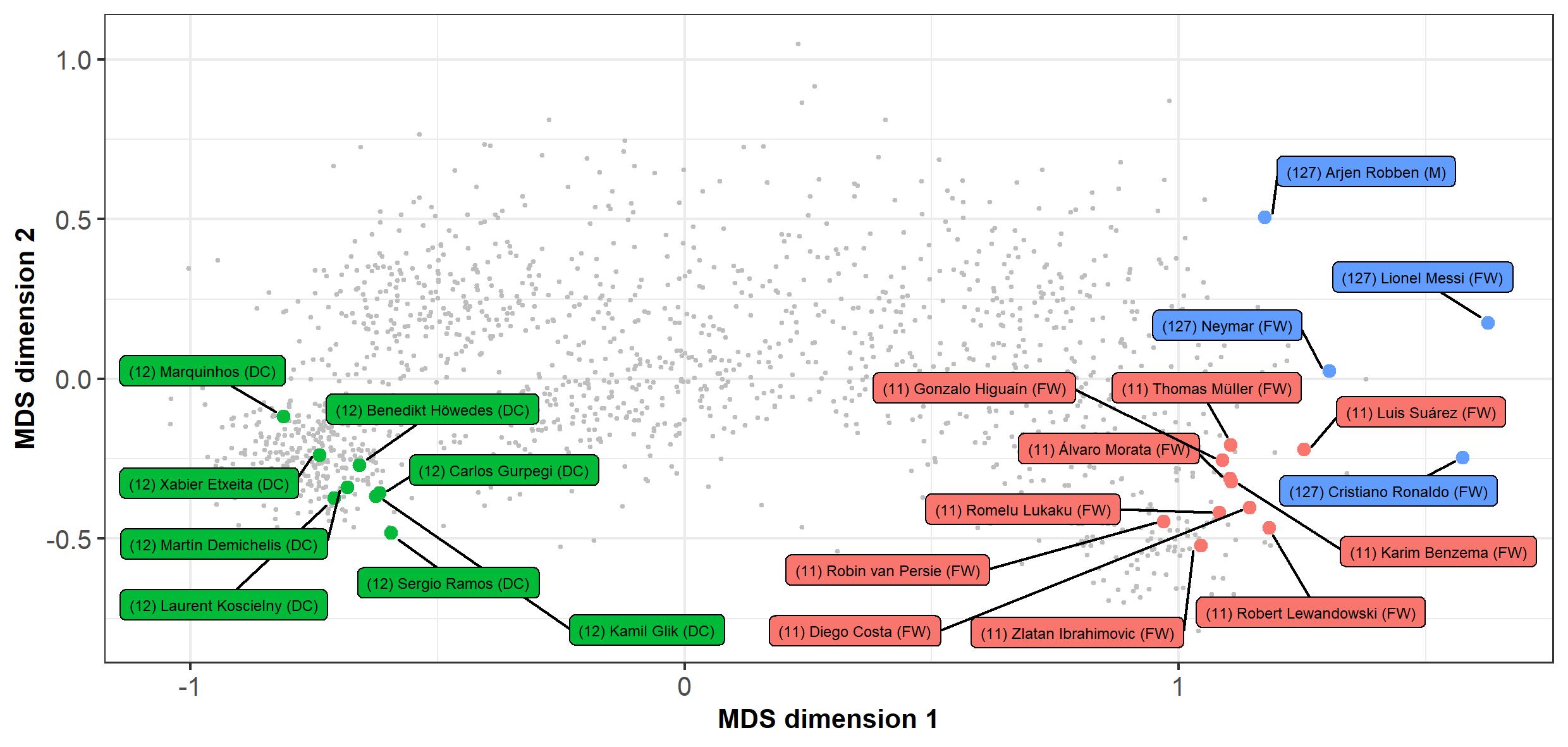}
  \caption{MDS plot of the football players data with three complete clusters of the PAM solution with $K=150$.}
  \label{fig:football_mds_pam_150}
\end{figure}

Interpreting all 150 clusters is infeasible here, so we focus in just three clusters, see Figure~\ref{fig:football_mds_pam_150}.
The most obvious result is that some of the most well known forward players (Messi, Ronaldo, Neymar and Robben) are grouped in one cluster, no. 127. These players are in Figure~\ref{fig:football_mds_pam_150} well distanced from the other players. They stand out especially in attacking features, such as shot, goal, dribble, key pass, but are also, atypically for general forward players, strong at short passes, see Table~\ref{tab:top_level_summary} in the Appendix. The PAM objective function allows to group them together despite a considerable within-cluster variance, which is better in terms of entropy than isolating them individually as ``outliers'', as happened in some other clusterings with large $K$.

Cluster 12 has typical central defenders who are skilled in variables such as clearance and aerial duels, while the players in cluster 11 are strikers who are well characterised by seemingly more negative aspects such as offsides, dispossession and bad control. Regarding positive characteristics, they are strong regarding shots and goals, but not as strong as cluster 127. Compared with cluster 127, they are stronger in aerial duels and clearances, but despite well reputed players being in this cluster, it can be clearly seen that they are not as outstanding as those in cluster 127.

Finding the optimal clustering at the largest considered number of clusters $K=150$ suggests that even better results may be achieved at even larger $K$. Ultimately we do not believe that any single clustering, particularly at such fine granularity, can be justified as the objectively best one. $K=150$ is probably large enough in practice, but in principle, accepting a high computational burden, the methodology can be extended to larger $K$.   

\subsection{Other indexes}
\label{sec:clustering_comparison}
On top of the results of $\mathcal{A}_1(\mathcal{C})$ and $\mathcal{A}_1(\mathcal{C})$, Table~\ref{tab:football_data_validitiy_index_comparison} also shows the best clusterings according to some validity indexes from the literature that are meant to measure the general quality of a clustering, as mentioned in Section \ref{sec:indexes}. The $K=2$ solutions for single linkage and spectral clustering marked as optimal by ASW, CH, Pearson$\Gamma$, and Bootstab, contain a very small cluster with outstanding players and do not differentiate between the vast majority of players. The complete linkage solution that is optimal according to Dunn's index belongs to Selection 7 that comes out worst in the survey of football experts, see Table \ref{tab:survey_res}. CVNN (run with tuning parameter $\kappa=10$, see \cite{liu2013understanding}) achieves best results for Ward's method with $K=4$ and $K=5$, which is reasonably in line with our $\mathcal{A}_1(\mathcal{C})$.

\begin{table}[tbp]
	\tiny
	\caption{Clustering validity index results for the football players data; note that for Bootstab and CVNN smaller values are better.}
		\begin{tabular}{l|ccccc}
		\multirow{2}{*}{\textbf{Validity Index}} & \multicolumn{5}{c}{\textbf{\underline{Best clusterings in order ($K$) with validity index values}}}\\
		& \textbf{First} & \textbf{Second} & \textbf{Third}& \textbf{Fourth} & \textbf{Fifth} \\[0.25em]
		\hline
         $\mathcal{A}_{1}(\mathcal{C})$ & $Ward \,(5)$ & $Ward \,(6)$ & $PAM \,(6)$ & $PAM \,(5)$ & $Ward \,(4)$ \\    
                     & 1.386 & 1.336 & 1.216 & 1.172 & 1.081 \\[0.25em]
         $\mathcal{A}_{2}(\mathcal{C})$ & $PAM \,(150)$ & $PAM \,(149)$ & $PAM \,(148)$ & $PAM \,(147)$ & $PAM \,(146)$ \\    
                     & 1.025 & 1.021 & 1.020 & 1.019 & 1.017 \\[0.25em]
                 \hline
 		$ASW$ & $Spectral \,(2)$ & $Average \,(2)$ & $Ward \,(2)$ & $PAM \,(2)$ & $Complete \,(2)$  \\
 		      & $0.345$         & $0.344$        & $0.342$     & $0.340$    & $0.340$          \\[0.25em]
 		$CH$  & $Spectral \,(2)$ & $PAM \,(2)$ & $Complete \,(2)$ & $Average \,(2)$ & $Ward \,(2)$ \\
 		      & $1038$      & $1027$ & $1013$      & $1006$      & $967$             \\[0.25em]
 	    $Dunn$ & $Complete \,(145)$ & $Complete \,(144)$ & $Complete \,(143)$ & $Complete \,(142)$ & $Complete \,(141)$ \\
 		       & $0.371$           & $0.371$           & $0.371$           & $0.370$           & $0.368$    \\[0.25em]	
         $Pearson \, \Gamma$ & $Spectral \,(2)$ & $Average \,(2)$ & $Ward \,(2)$ & $Average \,(4)$ & $Complete \,(2)$ \\
                             & $0.695$         & $0.693$        & $0.693$     & $0.692$        & $0.687$    \\[0.25em]
         $CVNN$ & $Ward \,(4)$ & $Ward \,(5)$ & $PAM \,(4)$ & $Ward \,(3)$ & $PAM \,(5)$ \\              
                             & $0.935$     & $0.965$     & $0.976$    & $0.988$     & $1.034$   \\[0.25em]
        $Bootstab$ & $Single \,(2)$ & $Single \,(3)$ & $Single \,(4)$ & $Single \,(5)$ & $PAM \,(150)$ \\[0.25em]   
 & $0.0011$ & $0.0021$ & $0.0025$  & $0.0039$ & $0.0039$ \\[0.25em]
		\end{tabular}
	\label{tab:football_data_validitiy_index_comparison}	
\end{table}


 
\section{Conclusion}
\label{sec:conclusion}
We computed two different clusterings of football player performance data from the 2014-15 season. We believe that the considerations presented here are worthwhile also for analysing new data, in particular regarding dissimilarity construction, measuring desirable characteristics of a clustering, and using such measurement to select a specific clustering. Results from the approach taken here look more convincing than the assessments given by existing indexes from the literature that attempt to quantify clustering quality in a one-dimensional manner. The index combination from calibrated average within-cluster dissimilarities, Pearson-$\Gamma$, entropy, Bootstab stability, and (with half the weight) separation may generally be good for balancing within-cluster homogeneity and ``natural'' separation as far as it occurs in the data in situations where for interpretative reasons useful clusters should have roughly the same size. The focus of this combination is a bit stronger on within-cluster homogeneity than on separation. Chances are that natural variation between human beings implies that athletes' performance data will not normally be characterised by strong separation between different groups, particularly not if such groups are not very homogeneous. The involvement of stability should make sure that the found clusters are not spurious. 

The second combination of indexes used here, Bootstab with full weight and entropy with half weight, was motivated by best agreement with football expert's assessments based on the specific data set analysed here. One may wonder whether this is a good combination also for different data for finding a clustering on a finer scale, i.e., with more and smaller clusters. Entropy is in all likelihood important for the use of such a clustering; endemic occurrence of clusters with one or two players should be avoided. Stability is certainly desirable in itself; it is also correlated over all involved clusterings strongly (0.629) with low average within-cluster dissimilarities, so it carries some information on within-cluster homogeneity, too. Strong between-cluster separation in absolute terms can hardly be expected with such a large number of clusters; these clusterings have a pragmatic use rather than referring to essential underlying differences between them. Although it is conceivable that this index combination works well also for new in some sense similar data, a wider investigation into which characteristics of clusterings correspond to expert assessments of their use and plausibility would surely be of interest.

The proposed methodology is implemented in the function clusterbenchstats in the R-package fpc \citep{fpc}.

\subsection*{Acknowledgments}
	We are very thankful to İstanbul Başakşehir Football Club to give the opportunity for making this survey and provided us a network with other football experts. such as journalists.  

\subsection*{Funding}
	The work of the second author was supported by EPSRC grant EP/K033972/1.





\begin{appendix}
\section{Basic statistical summary of top level count variables for the selected clusters}

{\fontsize{4}{4}\selectfont
\begin{table}[htbp]
	\renewcommand{\arraystretch}{1.25}
	\caption{Statistical summary (Mean $\pm$ standard deviation) of top level variables for no cluster solution (Overall), 3 cluster solutions in PAM ($K=150$) and all cluster solutions in Ward ($K=5$).  \label{tab:top_level_summary} }
	\begin{tabular}{l|l|lll|lllll}
		\hline
		\multirow{2}{*}{\textbf{Variables}} & \multirow{2}{*}{\textbf{Overall}} & \multicolumn{3}{c}{\underline{\textbf{PAM ($K=150$)}}} & \multicolumn{5}{c}{\underline{\textbf{Ward ($K=5$)}}}\\
		& & \textbf{Cluster 11} & \textbf{Cluster 12} & \textbf{Cluster 127} & \textbf{Cluster 1} & \textbf{Cluster 2} & \textbf{Cluster 3} & \textbf{Cluster 4} & \textbf{Cluster 5} \\
		\hline
TACKLE        &  1.96  $\pm$ 0.88 &  0.66  $\pm$ 0.28 & 1.82  $\pm$ 0.34 & 0.70   $\pm$ 0.39 & 2.62  $\pm$ 0.80 & 2.42  $\pm$ 0.69 &  1.80   $\pm$ 0.58 & 1.67  $\pm$ 0.65 & 0.77 $\pm$ 0.39 \\
INTERCEPTION  &  1.77  $\pm$ 0.99 &  0.32  $\pm$ 0.15 & 3.13  $\pm$ 0.58 & 0.29  $\pm$ 0.19 &  2.04  $\pm$ 0.74 & 2.28  $\pm$ 0.67 &  2.59  $\pm$ 0.74 & 0.96  $\pm$ 0.41 & 0.42 $\pm$ 0.25 \\
FOUL GIVEN    &  1.38  $\pm$ 0.73 &  1.36  $\pm$ 0.52 & 0.74  $\pm$ 0.39 & 2.27  $\pm$ 1.03 &  1.46  $\pm$ 0.67 & 1.17  $\pm$ 0.54 &  0.80   $\pm$ 0.38 & 1.88  $\pm$ 0.78 & 1.69 $\pm$ 0.68 \\
FOUL COMMITTED &  1.43  $\pm$ 0.61 &  1.42  $\pm$ 0.43 & 1.12  $\pm$ 0.48 & 0.86  $\pm$ 0.27 &  1.69  $\pm$ 0.61 & 1.29  $\pm$ 0.46 &  1.12  $\pm$ 0.42 & 1.42  $\pm$ 0.61 & 1.77 $\pm$ 0.77 \\
OFFSIDE       &  0.22  $\pm$ 0.34 &  1.13  $\pm$ 0.37 & 0.06  $\pm$ 0.07 & 0.67  $\pm$ 0.30  &  0.06  $\pm$ 0.07 & 0.08  $\pm$ 0.09 &  0.03  $\pm$ 0.05 & 0.33  $\pm$ 0.27 & 0.88 $\pm$ 0.38 \\
CLEARANCE     &  2.60   $\pm$ 2.43 &  0.77  $\pm$ 0.33 &    5.00  $\pm$ 0.93 & 0.29  $\pm$ 0.40  &  1.54  $\pm$ 0.80  & 3.11  $\pm$ 1.26 &  6.37  $\pm$ 1.77 & 0.63  $\pm$ 0.51 &  0.70 $\pm$ 0.45 \\
BLOCK         &  1.44  $\pm$ 0.55 &  0.48  $\pm$ 0.19 & 1.57  $\pm$ 0.25 & 0.75  $\pm$ 0.25 &  1.58  $\pm$ 0.48 & 1.82  $\pm$ 0.46 &  1.48  $\pm$ 0.39 & 1.32  $\pm$ 0.47 & 0.72 $\pm$ 0.34 \\
SHOT          &  1.29  $\pm$ 1.01 &  3.44  $\pm$ 0.48 & 0.47  $\pm$ 0.28 & 4.94  $\pm$ 1.39 &  1.09  $\pm$ 0.55 & 0.56  $\pm$ 0.38 &  0.50   $\pm$ 0.26 & 2.12  $\pm$ 0.78 & 2.76 $\pm$ 0.84 \\
GOAL          &  0.14  $\pm$ 0.17 &  0.62  $\pm$ 0.10  & 0.12  $\pm$ 0.06 &    1.00  $\pm$ 0.21 &  0.08  $\pm$ 0.08 & 0.03  $\pm$ 0.05 &  0.05  $\pm$ 0.05 & 0.23  $\pm$ 0.14 & 0.43 $\pm$ 0.20 \\
DRIBBLE       &  0.91  $\pm$ 0.81 &  1.42  $\pm$ 0.34 & 0.21  $\pm$ 0.11 & 3.99  $\pm$ 1.52 &  0.79  $\pm$ 0.60  &  0.80  $\pm$ 0.50  &  0.30   $\pm$ 0.25 & 1.70   $\pm$ 1.00 & 0.94 $\pm$ 0.75 \\
UNSTCH        &  1.16  $\pm$ 0.81 &  2.20   $\pm$ 0.66 & 0.33  $\pm$ 0.18 & 1.59  $\pm$ 0.41 &  0.93  $\pm$ 0.39 &  0.80  $\pm$ 0.32 &  0.32  $\pm$ 0.16 & 1.90   $\pm$ 0.58 & 2.28 $\pm$ 0.62 \\
DISPOSSESSED  &  1.15  $\pm$ 0.86 &  1.99  $\pm$ 0.68 & 0.19  $\pm$ 0.08 & 1.73  $\pm$ 0.39 &  1.12  $\pm$ 0.51 & 0.73  $\pm$ 0.39 &  0.23  $\pm$ 0.16 & 2.03  $\pm$ 0.68 & 1.96 $\pm$ 0.74 \\
AERIAL DUEL        &  1.80   $\pm$ 1.29 &  1.72  $\pm$ 0.73 & 2.55  $\pm$ 0.76 & 0.91  $\pm$ 0.60  &  1.51  $\pm$ 1.02 & 1.45  $\pm$ 0.84 &  2.84  $\pm$ 1.06 & 1.05  $\pm$ 1.00 & 2.48 $\pm$ 1.71 \\
PASS (SHORT)    &  37.0 $\pm$ 12.3 &  30.0 $\pm$ 6.10  & 46.7 $\pm$ 7.69 & 52.3 $\pm$ 10.6 &  47.8 $\pm$ 12.6& 34.5 $\pm$ 8.50 &  37.0 $\pm$ 11.7 & 35.0 $\pm$ 9.42 & 25.7 $\pm$ 7.14 \\
PASS (LONG)     &  5.13  $\pm$ 3.10  &  1.10   $\pm$ 0.45 & 5.91  $\pm$ 1.93 & 2.05  $\pm$ 1.20  &  5.96  $\pm$ 2.54 & 5.75  $\pm$ 1.99 &  8.28  $\pm$ 2.37 & 2.89  $\pm$ 1.57 & 1.14 $\pm$ 0.80 \\
CROSS         &  2.07  $\pm$ 2.19 &  1.29  $\pm$ 0.76 & 0.18  $\pm$ 0.20  & 2.54  $\pm$ 0.99 &  1.46  $\pm$ 1.66 & 3.05  $\pm$ 1.71 &  0.21  $\pm$ 0.45 & 4.09  $\pm$ 2.48 & 1.17 $\pm$ 0.99 \\
CORNER        &  0.54  $\pm$ 1.07 &  0.07  $\pm$ 0.10  &    0.00  $\pm$ 0.00    & 1.40   $\pm$ 1.51 &  0.61  $\pm$ 1.14 & 0.26  $\pm$ 0.68 &  0.01  $\pm$ 0.22 & 1.46  $\pm$ 1.36 & 0.14 $\pm$ 0.53 \\
FREE KICK      &  1.16  $\pm$ 0.98 &  0.13  $\pm$ 0.11 & 1.44  $\pm$ 0.53 & 0.71  $\pm$ 0.44 &  1.54  $\pm$ 1.16 & 1.17  $\pm$ 0.63 &  1.92  $\pm$ 0.79 & 0.63  $\pm$ 0.62  & 0.11 $\pm$ 0.20 \\
KEY PASS       &  0.99  $\pm$ 0.70  &  1.56  $\pm$ 0.36 & 0.17  $\pm$ 0.09 & 2.41  $\pm$ 0.45 &  1.01  $\pm$ 0.58 & 0.84  $\pm$ 0.43 &  0.23  $\pm$ 0.15 & 1.75  $\pm$ 0.61 & 1.17 $\pm$ 0.48 \\
ASSIST        &  0.11  $\pm$ 0.12 &  0.25  $\pm$ 0.13 & 0.01  $\pm$ 0.02 & 0.41  $\pm$ 0.16 &  0.09  $\pm$ 0.11 & 0.09  $\pm$ 0.09 &  0.02  $\pm$ 0.04 & 0.21  $\pm$ 0.13 & 0.14 $\pm$ 0.11 \\
		\hline
		\multicolumn{10}{l}{\tiny{*UNSTCH: Unsuccessful Touch (Bad control)}} \\ 
	\end{tabular}
	\label{tab:cluster_results}
\end{table}}

\section{Survey questions}
Tables \ref{tab:survey_question1}-\ref{tab:survey_question7} list the questions from the survey of preferences of football experts regarding the grouping of certain players. The corresponding clusterings and Selection numbers from Table \ref{tab:survey_res_cluster_selections} are also included, although these were not shown to the experts.

\begin{table}[htbp]
	\footnotesize
	\caption{Question 1: This group of players are centre-defenders. Please rank the following in order of appropriateness from 1 to 5 where 1 is the most appropriate to you and 5 is the least appropriate to you.}
	\centering
	\begin{tabular}{c | c c c c |c}
	\thickhline 
	\textbf{Clustering solutions} & \textbf{1. Group} & \textbf{2. Group} & \textbf{3. Group} & \textbf{4. Group} & \textbf{Rank} \\
        \hline
	\shortstack[c]{\underline{Selection 1, 2, 3, 4} \\ PAM \\ ($K \sim [100:150]$)} & John Terry & Gary Cahill & Chris Smalling & \shortstack[c] {John Stones \\ Thiago Silva} & \\
	\hline
	\shortstack[c]{\underline{Selection 5} \\ Ward's method \\ ($K \sim [100:147] $)} & \shortstack[c]{John Terry \\ Gary Cahill \\ John Stones \\ Thiago Silva} & Chris Smalling & --- & --- & \\
	\hline
	\shortstack[c]{\underline{Selection 6} \\ Ward's method \\ ($K \sim [148:150] $)} & \shortstack[c]{John Terry \\ Gary Cahill \\ Thiago Silva} & Chris Smalling & John Stones & --- & \\
	\hline
	\shortstack[c]{\underline{Selection 7} \\ Complete linkage \\ ($K \sim [100:150] $)} & \shortstack[c]{John Terry \\ Gary Cahill \\ Thiago Silva} & \shortstack[c]{ Chris Smalling \\ John Stones} & --- & --- & \\
	\hline
	\shortstack[c]{\underline{Selection 8} \\ Average linkage \\ ($K \sim [100:150] $)} & \shortstack[c]{John Terry \\ Gary Cahill} & \shortstack[c]{Thiago Silva \\ Chris Smalling \\ John Stones} & --- & --- & \\
	\thickhline 
	\end{tabular}
	\label{tab:survey_question1}	
\end{table}

\begin{table}[htbp]
	\footnotesize
	\caption{Question 2: This group of players are right or left defenders. Please rank the following in order of appropriateness from 1 to 2 where 1 is the most appropriate to you and 2 is the least appropriate to you.}
	\centering
	\begin{tabular}{c | c c c |c}
	\thickhline 
	\textbf{Clustering solutions} & \textbf{1. Group} & \textbf{2. Group} & \textbf{3. Group} & \textbf{Rank} \\
        \hline
	\shortstack[c]{\underline{Selection 1, 2, 3, 4, 5, 6} \\ PAM and Ward's method \\ ($K \sim [100:150]$)} & Cesar Azpilicueta & Gael Clichy & \shortstack[c] {Dani Alves\\ Daniel Carvajal} & \\
	\hline
	\shortstack[c]{\underline{Selection 7,8} \\ Complete and average linkage \\ ($K \sim [100:150] $)} & \shortstack[c]{Cesar Azpilicueta \\ Gael Clichy} & \shortstack[c]{Dani Alves\\ Daniel Carvajal} & --- & \\
	\thickhline 
	\end{tabular}
	\label{tab:survey_question2}	
\end{table}

\begin{table}[htbp]
	\footnotesize
	\caption{Question 3: This group of players are defensive midfileders. Please rank the following in order of appropriateness from 1 to 3 where 1 is the most appropriate to you and 3 is the least appropriate to you.}
	\centering
	\begin{tabular}{c | c c c |c}
	\thickhline 
	\textbf{Clustering solutions} & \textbf{1. Group} & \textbf{2. Group} & \textbf{3. Group} & \textbf{Rank} \\
        \hline
	\shortstack[c]{\underline{Selection 1, 4} \\ PAM \\ ($K \sim [100:113, 130:133, 137:146]$)} & \shortstack[c]{Nemanja Matic \\ Fernando} & \shortstack[c] {Sergio Busquets \\ Javier Mascherano} & --- & \\
	\hline
	\shortstack[c]{\underline{Selection 5, 6} \\ Ward's method \\ ($K \sim [100:150]$)} & Nemanja Matic & Fernando & \shortstack[c] {Sergio Busquets \\ Javier Mascherano} & \\
	\hline
	\shortstack[c]{\underline{Selection 2, 3, 7, 8} \\ PAM \\ ($K \sim [114:129, 134:136, 147:150]$), \\ Complete and average linkage \\ ($K \sim [100:150]$)} & Nemanja Matic & \shortstack[c] {Fernando \\ Sergio Busquets \\ Javier Mascherano} & --- & \\
	\thickhline 
	\end{tabular}
	\label{tab:survey_question3}	
\end{table}

\begin{table}[htbp]
	\footnotesize
	\caption{Question 4: This group of players are midfileders. Please rank the following in order of appropriateness from 1 to 3 where 1 is the most appropriate to you and 3 is the least appropriate to you.}
	\centering
	\begin{tabular}{c | c c c c |c}
	\thickhline 
	\textbf{Clustering solutions} & \textbf{1. Group} & \textbf{2. Group} & \textbf{3. Group} & \textbf{4. Group} & \textbf{Rank} \\
        \hline
	\shortstack[c]{\underline{Selection 1, 4} \\ PAM \\ ($K \sim [100:113, 130:133, 137:146]$)} & Gabi & Tiago & Xabi Alonso & Thiago Motta & \\
	\hline
	\shortstack[c]{\underline{Selection 2, 3} \\ PAM \\ ($K \sim [114:129, 134:136, 147:150]$)} & Gabi & Tiago & \shortstack[c]{ Xabi Alonso \\ Thiago Motta} & --- & \\
	\hline
	\shortstack[c]{\underline{Selection 5, 6, 7, 8} \\ Ward's method, complete and average linkage \\ ($K \sim [100:150]$)} & \shortstack[c]{ Gabi \\ Xabi Alonso \\ Thiago Motta} & Tiago & --- & --- & \\
	\thickhline 
	\end{tabular}
	\label{tab:survey_question4}	
\end{table}

\begin{table}[htbp]
	\footnotesize
	\caption{Question 5: This group of players are defensive midfileders. Please rank the following in order of appropriateness from 1 to 3 where 1 is the most appropriate to you and 3 is the least appropriate to you.}
	\centering
	\begin{tabular}{c | c c c c |c}
	\thickhline 
	\textbf{Clustering solutions} & \textbf{1. Group} & \textbf{2. Group} & \textbf{3. Group} & \textbf{4. Group} & \textbf{Rank} \\
        \hline
	\shortstack[c]{\underline{Selection 1, 2, 3, 4, 8} \\ PAM and average linkage \\ ($K \sim [100:150]$)} & \shortstack[c]{ Paul Pogba \\ Arturo Vidal} & Kevin De Bruyne & Henrikh Mkhitaryan &  --- & \\
	\hline
	\shortstack[c]{\underline{Selection 5, 6} \\ Ward's method \\ ($K \sim [100:150]$)} & Paul Pogba & Arturo Vidal & Kevin De Bruyne & Henrikh Mkhitaryan & \\
	\hline
	\shortstack[c]{\underline{Selection 7} \\ Complete linkage \\ ($K \sim [100:150]$)} & \shortstack[c]{ Paul Pogba \\ Arturo Vidal} & \shortstack[c]{ Kevin De Bruyne \\ Henrikh Mkhitaryan} & --- & --- & \\
	\thickhline 
	\end{tabular}
	\label{tab:survey_question5}	
\end{table}

\begin{table}[htbp]
	\footnotesize
	\caption{Question 6: This group of players are attacking midfileders. Please rank the following in order of appropriateness from 1 to 5 where 1 is the most appropriate to you and 5 is the least appropriate to you.}
	\centering
	\begin{tabular}{c | c c c |c}
	\thickhline 
	\textbf{Clustering solutions} & \textbf{1. Group} & \textbf{2. Group} & \textbf{3. Group} & \textbf{Rank} \\
        \hline
	\shortstack[c]{\underline{Selection 1, 2} \\ PAM \\ ($K \sim [100:118]$)} & \shortstack[c]{Lionel Messi \\ Neymar \\ Arjen Robben} & Eden Hazard & Cristiano Ronaldo & \\
	\hline
	\shortstack[c]{\underline{Selection 3, 4} \\ PAM \\ ($K \sim [119:150]$)} & \shortstack[c]{Lionel Messi \\ Neymar \\ Arjen Robben \\ Cristiano Ronaldo} & Eden Hazard & --- & \\
	\hline
	\shortstack[c]{\underline{Selection 5, 6} \\ Ward's method \\ ($K \sim [100:150]$)} & \shortstack[c]{Lionel Messi \\ Arjen Robben \\ Cristiano Ronaldo} & \shortstack[c]{Eden Hazard \\ Neymar} & --- & \\
	\hline
	\shortstack[c]{\underline{Selection 7} \\ Complete linkage \\ ($K \sim [100:150]$)} & \shortstack[c]{Lionel Messi \\ Arjen Robben \\ Eden Hazard \\ Neymar} & Cristiano Ronaldo & --- & \\
	\hline
	\shortstack[c]{\underline{Selection 8} \\ Average linkage \\ ($K \sim [100:150]$)} & \shortstack[c]{Lionel Messi \\ Eden Hazard \\ Neymar} & Cristiano Ronaldo & Arjen Robben & \\
	\thickhline 
	\end{tabular}
	\label{tab:survey_question6}	
\end{table}

\begin{table}[htbp]
	\footnotesize
	\caption{Question 7: This group of players are forwards. Please rank the following in order of appropriateness from 1 to 5 where 1 is the most appropriate to you and 5 is the least appropriate to you.}
	\centering
	\begin{tabular}{c | c c c c |c}
	\thickhline 
	\textbf{Clustering solutions} & \textbf{1. Group} & \textbf{2. Group} & \textbf{3. Group} & \textbf{4. Group} & \textbf{Rank} \\
        \hline
	\shortstack[c]{\underline{Selection 1, 2} \\ PAM \\ ($K \sim [100:118]$)} & \shortstack[c]{Cristiano Ronaldo \\ Karim Benzema} & Robert Lewandowski & Zlatan Ibrahimovic & --- & \\
	\hline
	\shortstack[c]{\underline{Selection 3, 4} \\ PAM \\ ($K \sim [119:150]$)} & Cristiano Ronaldo & \shortstack[c]{Robert Lewandowski \\ Zlatan Ibrahimovic} & Karim Benzema & --- & \\
	\hline
	\shortstack[c]{\underline{Selection 5, 6} \\ Ward's method \\ ($K \sim [100:150]$)} & Cristiano Ronaldo & \shortstack[c]{Robert Lewandowski \\ Zlatan Ibrahimovic \\ Karim Benzema} & --- & --- & \\
	\hline
	\shortstack[c]{\underline{Selection 7} \\ Complete linkage \\ ($K \sim [100:150]$)} & \shortstack[c]{ Cristiano Ronaldo \\ Karim Benzema} & \shortstack[c]{Robert Lewandowski \\ Zlatan Ibrahimovic} & --- & --- & \\
	\hline
	\shortstack[c]{\underline{Selection 8} \\ Average linkage \\ ($K \sim [100:150]$)} & Cristiano Ronaldo & Karim Benzema & Robert Lewandowski & Zlatan Ibrahimovic & \\
	\thickhline 
	\end{tabular}
	\label{tab:survey_question7}	
\end{table}

\end{appendix}

\end{document}